\documentclass[onecolumn,a4paper,showabstract]{revtex4-2}
\setcitestyle{super}

\usepackage[utf8]{inputenc}
\usepackage{amsmath,amsfonts,amssymb}
\usepackage{graphicx,graphics}
\usepackage{epstopdf,csquotes,}
\usepackage{color,textcomp,bm}
\usepackage{subfigure,array,xspace} 
\usepackage[squaren,Gray,cdot]{SIunits}
\usepackage{float}
\usepackage[pdfstartview=FitV,bookmarks=true, linktocpage=true,colorlinks=true,
            urlcolor=blue,linkcolor=red, citecolor=blue]{hyperref}   
\usepackage{mathdots}
\usepackage{MnSymbol}
\usepackage{placeins}
\usepackage{multibib}

\newcommand{\alex}[1]{\textcolor{black}{#1}}
\newcommand{\corr}[1]{\textcolor{black}{#1}}
\newcommand{\al}[1]{\textcolor{black}{#1}}
\newcommand{\fab}[1]{\textcolor{black}{#1}}
\newcommand{\rev}[1]{\textcolor{black}{#1}}
\newcommand{\re}[1]{\textcolor{black}{#1}}
\newcommand{\ree}[1]{\textcolor{black}{#1}}

\graphicspath{{Figures/}}
\begin{abstract}
\vspace{5 mm}

{\noindent {\large \textbf{Abstract}}} \\

We report on experimental and numerical implementations of devices based on the negative refraction of elastic guided waves, the so-called Lamb waves. Consisting in plates of varying thickness, these devices rely on the concept of complementary media, where a particular layout of negative index media can cloak an object with its anti-object or trap waves around a negative corner. The diffraction cancellation operated by negative refraction is investigated by means of laser ultrasound experiments. However, unlike original theoretical predictions, these intriguing wave phenomena remain, nevertheless, limited to the propagating component of the wave-field. To go beyond the diffraction limit, negative refraction is combined with the concept of metalens, a device converting the evanescent components of an object into propagating waves. The transport of an evanescent wave-field is then possible from an object plane to a far-field imaging plane. Twenty years after Pendry’s initial proposal, this work thus paves the way towards an elastic superlens. 
\end{abstract}

\begin{document}

\title{\re{Cloaking, Trapping and Superlensing of Lamb Waves with Negative Refraction}}
\author{François Legrand{$^\dag$}}
\author{Benoît Gérardin{$^\dag$}}
\author{François Bruno} 
\author{Jérôme~Laurent}
\author{Fabrice Lemoult}
\author{Claire Prada} 
\author{Alexandre Aubry{$^*$}} 
\affiliation{ \vspace{5mm} Institut Langevin, ESPCI Paris, PSL University, CNRS, 1 rue Jussieu, F-75005 Paris, France \vspace{5mm}\\
{$^\dag$}These authors equally contributed to this work\\
{$^*$}Corresponding author (e-mail: alexandre.aubry@espci.fr)\vspace{5mm}}


\maketitle

\noindent {\large \textbf{Introduction}} \\

For the last 20 years, negative refraction has received a considerable attention for its ability of cancelling wave diffraction~\cite{veselago1968,pendry2000negative,pendrycomp,leonhardt2006optical}. Any negative refracting slab constitutes a flat lens which does not suffer from any spherical aberration. Better yet, it actually forms a perfect lens able to resolve details much smaller than the conventional diffraction limit in wave imaging~\cite{pendry2000negative,Fang2005}. More generally, wave diffraction can be cancelled by adjoining two complementary media, \textit{i.e} two mirror regions of opposite refractive indices~\cite{pendrycomp,leonhardt2006optical}.

These notions of perfect lens and complementary media have drawn considerable attention in the physics community. Yet, the experimental implementations of those concepts have remained relatively limited so far. \fab{Many attempts have been proposed~\cite{liu2003rapid,Hu2004,Fang2005,taubner2006near,zhang2008superlenses,Sukhovich2009,zhu2014negative,Park2015,Brunet2015,Kaina2015,xu2017realization} to build artificial media, such as photonic/phononic crystals or metamaterials, which can be described by a negative index material. \ree{However,} these man made media first suffer from the inherent periodicity which imposes a limitation on the negative refracting lens resolution~\cite{Smith2003}. Also, the intrinsic resonant nature of many of the designs induces strong energy dissipation losses, which limit the depth-of-field and the resolving power of the devices. }All these features, in addition to the manufacturing imperfections, explain \re{the fact that most of the implemented superlenses based on negative refraction are only efficient in the near-field and fail in transporting the evanescent component of an object in the far-field.} 

More recently, an alternative route has been proposed for negative refraction. It consists in taking advantage of the complex dispersion properties exhibited by the guided elastic waves supported by a plate, the so-called Lamb modes. When two dispersion branches show close cut-off frequencies corresponding to a longitudinal and a transverse thickness mode of the same symmetry, their repulsion gives rise to a backward mode~\cite{tolstoy1957wave}. This negative phase velocity branch displays a minimum corresponding to a zero-group velocity (ZGV) point~\cite{prada2005laser,holland2003air,Xie2019}. \fab{At slightly higher frequency, exploiting the existence of a negative phase velocity mode,} negative refraction of Lamb waves has been achieved without any metamaterial\fab{. The underlying mechanism consists in a} mode conversion between forward and backward propagating modes (or vice versa) \fab{either} at a step-like thickness discontinuity~\cite{bramhavar2011negative,philippe2015focusing} \al{or at the interface between two different materials with an adequate acoustic impedance mismatch~\cite{Manjunath2019}.} More recent studies investigated the negative reflection of Lamb waves at a free plate edge~\cite{gerardin2016negative,veres2016broad,gerardin2019negative} and the conversion of propagating modes at a thickness step in order to optimize the negative refraction effect~\cite{legrand2018negative}. 

\begin{figure*}[htbp]
  \centering
  \includegraphics[width=\textwidth]{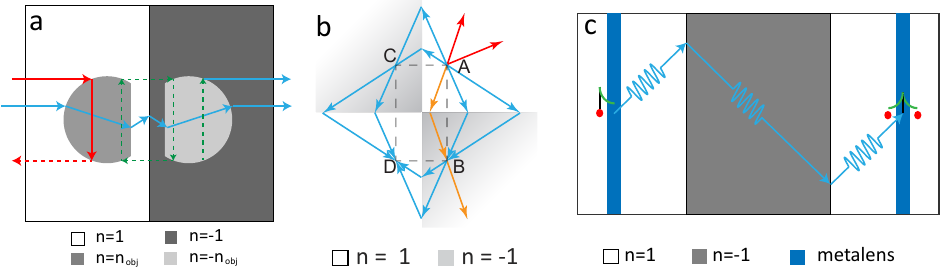}
  \caption{\alex{\textbf{Taming waves with complementary media}. \al{\textbf{a} Cloaking by an anti-object: Two slabs of equal thickness and placed adjacent to one another cancel wave diffraction if both display anti-symmetric refractive index distributions with respect to their common interface. Based on this principle, any object can be made invisible by its complementary counterpart, the so-called anti-object. \textbf{b} Wave trapping by a double perfect corner: A subset of rays diverging from a point source is returned to the source by means of successive negative refraction events. These waves then continue to circulate around the corner before being eventually absorbed. \textbf{c} The combination of a negative refraction slab with metalenses yields a superlens: (\textit{i}) The evanescent components of a point source is converted into propagating waves by interacting with \rev{an arrangement of sub-wavelength scatterers, the so-called metalens}; (\textit{ii}) This propagating wave-field is replicated in a far-field imaging plane by means of a negative refracting lens; (\textit{iii}) Introducing a second metalens, \rev{centrally symmetric} to the first one, in the imaging plane back-converts these propagating waves \rev{into the original evanescent components of the source by virtue of spatial reciprocity. A sub-wavelength image of the source is finally obtained.}}}}\label{fig:SchemaCroissantAndPC}
\end{figure*}

Building on these previous works, we investigate the concept of complementary media through the realisation of two devices \fab{consisting in duralumin plates of varying thickness}. \ree{The first one exploits the idea of anti-object in order to cloak a region of interest~\cite{lai2009complementary,Nguyen,Nguyen2019}.} This is done by adjoining a mirror region of \al{opposite index} that cancels the diffraction undergone by the waves inside the first region (see Fig.~ \ref{fig:SchemaCroissantAndPC}a). In a second step, we show how such a complementary  medium can behave as a trap if the waves are generated inside it ~\cite{notomi2002negative,pendry2003focusing}. Diffraction cancellation via negative refraction makes the wave circulate for ever around negative corners (see Fig.~\ref{fig:SchemaCroissantAndPC}b).
In this work, the design of both devices is based on a semi-analytical model~\cite{legrand2018negative} and optimized thanks to \al{finite difference time-domain (FDTD)} simulations~\cite{bossy2004three,simsonic}. \fab{The propagation of Lamb waves across such plates accordingly is
then investigated experimentally by means of laser interferometry.} 
On the one hand, diffraction cancellation is assessed by the Strehl ratio~\cite{mahajan1982strehl}, a parameter that quantifies wave distortions at the output of the complementary cloak. \al{On the other hand,} wave circulation around negative corners is highlighted by investigating the time-dependence of the wave-field. 
However, unlike Pendry's initial proposal~\cite{pendrycomp}, only the propagating component of the wave-field is properly recombined in each device. \ree{Indeed, the underlying mechanism for near-field cloaking and superlensing, the so-called anomalous resonance~\cite{Milton2005,milton2006cloaking,Nguyen,McPhedran2020}, requires more stringent conditions than a mere anti-symmetric distribution of the phase velocity~\cite{McPhedran2020,Deng2020}}. 

\re{To circumvent this issue, the third system under study aims at 
coupling the negative refraction phenomenon with the concept of metalens.} \rev{As shown by} Lemoult et al. ten years ago~\cite{lemoult2010resonant,lemoult2011acoustic,Lemoult2012}, an arrangement of sub-wavelength \rev{scatterers} can act as an efficient converter \rev{between the evanescent and propagating components} of a \rev{wave-field}. By placing such a metalens in the vicinity of both the object and imaging planes of the negative refracting lens (see Fig.\ref{fig:SchemaCroissantAndPC}c), the evanescent \rev{components} \rev{of the object} (or at least a part of it) can be transported from the object to the imaging plane. Again, our semi-analytical model and a numerical simulation enable the design of such an elastic superlens. The evanescent wave-field gives rise to a sub-wavelength focal spot of $\lambda/6$ at the device output. 

\vspace{5 mm}

\noindent {\large \textbf{Results}} \\

\noindent \textbf{\alex{Harnessing forward and backward Lamb modes.}}

Elastic waves consist in two - longitudinal and transverse - bulk acoustic waves traveling at distinct but unique velocities for a given material. In a plate, those elastic waves couple to each other at each reflection on the edges of the plate, giving rise to two infinite sets of dispersive Lamb modes. The particle motion of these modes lies in the sagittal plane that contains the direction of wave propagation and the normal of the plate. Lamb modes induce a deformation that can either be symmetric ($S_i$) or antisymmetric ($A_i$) with respect to the median plane. 

\begin{figure*}[htbp]
    \centering
    \includegraphics[width=0.6\textwidth]{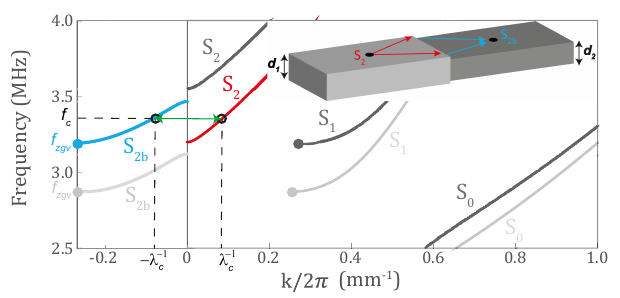}
    \caption{\alex{\textbf{Forward and backward Lamb modes}.} Dispersion curves \alex{of the symmetric Lamb modes in duralumin plates of $d_1=1$ mm (light grey) and $d_2=0.9$ mm thickness (dark grey). \al{Among all the symmetric modes supported by plate, the forward $S_2$ mode (red) in the thick part and the backward mode $S_{2b}$ (blue) in the thin part are of special interest for negative refraction. Their phase velocity coincide at crossing frequency $f_c$.}}}
    \label{fig:CdispDural}
\end{figure*}

In this article, the propagation of elastic waves across a duralumin plate (aluminium alloy) is studied. The material density is $\al{\rho}= 2795$ kg.m$^{-3}$. \alex{Its} longitudinal and \alex{shear} wave velocities are $c_L = 6400$ m.s$^{-1}$ and $c_T = 3120$ m.s$^{-1}$\alex{, respectively}. The dispersion curves of symmetric Lamb modes in a 1 mm\alex{-}thick \alex{plate} are displayed in green in Fig.~\ref{fig:CdispDural}.

The symmetric zero-order mode $S_0$ is the fundamental extensional mode of the plate that exhibits free propagation to zero frequency. On the contrary, the higher-order modes admit a cut-off frequency. One striking property of Duralumin is the small gap between the cut-off frequencies of the first- and second-order symmetric Lamb modes, $S_1$ and $S_2$. The repulsion between these two modes gives rise to a backward Lamb mode, referred to as $S_{2b}$, that coexists with the forward mode $S_1$~\cite{meitzler1965backward} just above the ZGV point ($f_{ZGV}$, $k_{ZGV}$)~\cite{prada2005laser,holland2003air} (see Fig.~\ref{fig:CdispDural}). \re{Note that other ZGV points can result from the repulsion of higher-order modes (for instance S3/S6 and S5/S10 in Duralumin~\cite{Prada2008}). Here we will only exploit the first ZGV point displayed by Fig.~\ref{fig:CdispDural} but similar negative refraction phenomena can be obtained from ZGV points at larger frequencies.}

\fab{If one considers a single Lamb mode impinging at an interface between two plates of different thickness, one can expect a large number of reflected and transmitted modes which can either be propagative, inhomogeneous  or evanescent~\cite{legrand2018negative}}
\fab{One} \alex{striking} phenomenon is that, at the crossing point between the forward mode in the thickest part (\al{red line} in Fig.~\ref{fig:CdispDural}) and the backward mode in the thinnest part (\al{blue line} in Fig.~\ref{fig:CdispDural}), the $S_2$ incident mode is mainly transmitted into the $S_{2b}$ mode. \fab{The thickness step should be made symmetric with respect to the mid-plane of the plate in order to avoid conversion into anti-symmetric Lamb modes.}
A semi-analytical study of this phenomenon~\cite{legrand2018negative} has allowed \alex{the optimization of the transmission coefficient between the $S_2$ and $S_{2b}$ modes.} \al{ In such a geometry, the optimal thickness step ratio is close to $0.9$. The corresponding amplitude transmission coefficient is $T=0.93$.} Furthermore, this high transmission is effective over a large angular spectrum ($T>0.8$ over 60$^{\circ}$). 

This phenomenon arises from the equality of $S_2$ and $S_{2b}$ absolute wave numbers at the crossing point. This implies that these two modes are associated with similar stress-displacement fields and only differ by their opposite
wave vectors. In consequence, \rev{the impedance mismatch between forward and backward Lamb modes at a thickness step remains limited.} In this paper, we will show this peculiar property holds in more complex geometries and how \al{to harness negative refraction} to compensate the diffraction or trap the waves through the use of the complementary media concept.

\vspace{10 mm}

\noindent \textbf{\alex{Cloaking by an anti-object.}}
\alex{For a complex slab made of alternatively positive and negative index areas, the juxtaposition of a complementary \al{mirror slab with an \alex{opposite refractive} index distribution} allows the cancellation of wave diffraction~\cite{pendry2003focusing}}. This concept may be extended to various kinds of index distributions. \alex{For instance, the scattered wave-field induced by an object can be hidden by adjoining a so-called ``anti-object"~\cite{pendrycomp} as illustrated by Fig.~\ref{fig:SchemaCroissantAndPC}a. The anti-object is the negative mirror image of the object, the mirror being taken to lie on the interface between the two complementary slabs. The object is thus hidden for one observer downstream to the anti-object as the wave-fields at the input and output of the complementary slabs shall be identical.}

\alex{Our goal is here to implement this idea for guided elastic waves.} The \al{system} studied here is a 1 mm\alex{-}thick plate excited by a line source that emits the forward $S_2$ mode as a plane wave. The \al{designed} object is a $7.5$ mm-wide truncated disc \al{of $0.9$ mm thickness and $104$ mm diameter}. \alex{At the boundary of this scatterer, the forward $S_2$ mode is converted into the backward $S_{2b}$ mode.}
\begin{figure*}[tb]
  \centering
  \includegraphics[width=0.8\textwidth]{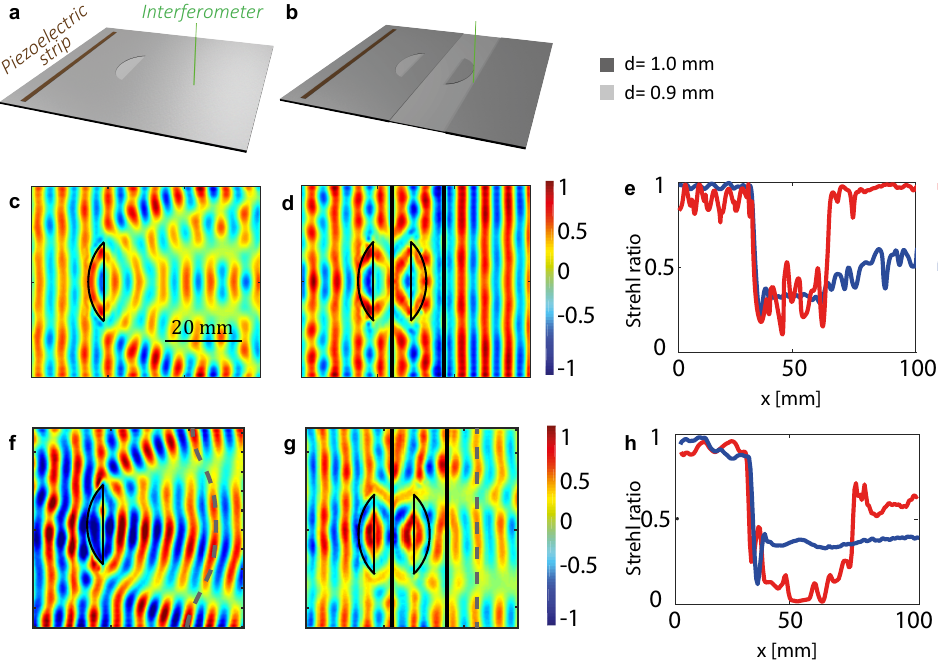}
  \caption{\alex{\textbf{Cloaking by an anti-object.} \textbf{a}, \textbf{b} \al{Experimental configuration} for the reference and complementary plates, respectively. \textbf{c}-\textbf{e} FDTD numerical simulation \re{(SimSonic3D software~\cite{bossy2004three}, www.simsonic.fr)}. The real part of the wave-field obtained in the reference (\textbf{c}) and complementary (\textbf{d}) plates is displayed at the crossing frequency \al{$f_c$ (see Fig.~\ref{fig:CdispDural})}. \textbf{e} Evolution of the Strehl ratio $\mathcal{S}(x)$ (Eq.~\ref{eq:Strehlratio}) across the reference (blue) and complementary (red) plate.  \textbf{f}-\textbf{h} Experimental result. The real part of the wave-field obtained in the reference (\textbf{f}) and complementary (\textbf{g}) plates is displayed at the crossing frequency \al{$f_c$}. \textbf{h} Evolution of $\mathcal{S}(x)$ across the reference (blue) and complementary (red) plate.}}\label{fig:croissant}
\end{figure*}
\alex{For the proof-of-concept, \al{two plate configurations are considered : the reference one with the object [see Fig.~\ref{fig:croissant}a] and the other one with the object and its anti-object [see Fig.~\ref{fig:croissant}b].} } 
 
\alex{These devices are first investigated numerically using a FDTD code~\cite{bossy2004three,simsonic}. The simulation parameters are described in the Methods section. A normal displacement pulse is applied to the line source. The normal displacement induced at the plate surface is recorded over a time length $\Delta t=110$ $\mu$s. A spatio-temporal filter described in the Methods section is then applied to the recorded wave-field in order to isolate the $S_2$ and $S_{2b}$ \al{mode contributions}.} 

\alex{Figure~\ref{fig:croissant}c shows the corresponding wave field in the reference plate at the crossing frequency $f_c=3.32$ MHz. \al{The corresponding wavelength $\lambda_c$ is of 12.2 mm.}} \al{Interferences between the incident and scattered waves result in phase and amplitude distortions, thus revealing the object to a downstream observer.} \alex{To quantify the impact of the object on the phase of the recorded wave-field, a relevant parameter is the Strehl ratio $\mathcal{S}$ ~\cite{mahajan1982strehl}. Often used in adaptive optics, $\mathcal{S}$ quantifies the phase distortions of a wave-front. In the experimental configurations displayed in Fig.~ \ref{fig:croissant}a and b, it can be defined as follows:
\begin{equation}
\label{eq:Strehlratio}
    \mathcal{S}(x) = \left |< e^{i \phi(x,y)} >_y\right |,
\end{equation}
where $\phi(x,y)$ is the phase of the wave field at the surface of the plate and the symbol \al{$\langle \cdots \rangle_y$ stands for a spatial average  along the $y-$direction}. The parameter $\mathcal{S}$ is equal to \al{unity} if the wave-field is a plane wave propagating along the $x-$direction, hence coinciding perfectly with the incident wave-field. On the contrary, the parameter $\mathcal{S}$ tends towards zero for a fully incoherent wave-field.}
 
\alex{Figure~\ref{fig:croissant}e displays the evolution of the \al{Strehl ratio} computed from the wave-field supported by the reference plate (Fig.~\ref{fig:croissant}c). Upstream to the object, the incident wave-field is not impacted ($\mathcal{S}\sim 0.98$) meaning that the object \al{does not induce any significant back scattering}. Its transverse cross-section is indeed larger than the \fab{wavelength}. The incident wave-field is thus scattered in the forward direction. \al{This} is quantified by the abrupt decay of the parameter $\mathcal{S}$ from 0.98 to 0.2 at the object's boundary. After the object, the wave-field remains distorted (Fig.~\ref{fig:croissant}c) with a Strehl ratio $\mathcal{S}$ remaining below $0.5$ (Fig.~\ref{fig:croissant}e).}

\alex{The addition of a complementary slab including the anti-object can compensate for these strong wave-distortions (Fig.~\ref{fig:croissant}b]). \al{The displacement field is calculated for the complementary plate using the FDTD simulation and displayed in Fig.~\ref{fig:croissant}d at the crossing frequency $f_c=3.32$ MHz.} The phase distortions accumulated by the wave through the object's band are now perfectly compensated by the anti-object's band. This phenomenon is induced by the conversion between the forward and backward modes at the interface between the two bands. The wave-front at the output of the complementary band retrieve the shape of the incident wave-front at the input, as if the object and complementary bands had disappeared. The anti-object thus enables the cloaking of the object to a downstream observer.}

\alex{The evolution of the Strehl ratio across the complementary plate provides a quantification of the cloaking performance (red curve in Fig.~\ref{fig:croissant}e). Upstream to the object, $\mathcal{S}$ fluctuates around a value of 0.9. These oscillations are a manifestation of the spurious reflections arising at each thickness step. As mentioned \al{previously}, the conversion between the $S_2$ and $S_{2b}$ modes is not perfect and residual back-reflections take place at each interface. The parameter $\mathcal{S}$ then abruptly decays in the complementary slabs to reach a value of 0.3, before retrieving a close to ideal value ($\mathcal{S}\sim 0.98$) at the device output. This excellent \al{Strehl ratio} confirms that the object is efficiently cloaked by its anti-object in transmission, while spurious reflections prevent from a perfect cloaking in reflection. }

Going further, it is interesting to notice that the theoretical \rev{amplitude} transmission coefficient at each thickness step is $\rev{T^2} \sim \rev{0.85}$~\cite{legrand2018negative} and thus the global transmission coefficients through the \al{six} steps crossed by the wave should be $\rev{T^{12}} \sim \rev{0.38}$. Yet, one can observe that the transmission through the devices is remarkably good. This striking observation comes from the elegant physics of complementary media: the reflections at the interfaces in the first band interfere destructively with the same reflections in the complementary band~\cite{pendrycomp}.

Following this numerical study, the devices displayed in Fig.~\ref{fig:croissant}a and b have been fabricated using a 1 mm-thick plate whose negative index areas have been etched by means of several engraving techniques described in the Methods section. The thickness map of each plate is provided in Supplementary Fig.~S1.
A piezoelectric strip is glued on the thick part of the plate to generate an incident plane wave (forward $S_2$ mode). The normal displacement at the plate surface is measured by means of a photorefractive interferometer (see Methods section).

Figures \ref{fig:croissant}f and g show the corresponding wave-fields at the crossing frequencies in the reference and complementary plates, respectively. \alex{These wave-fields show some difference compared to their numerical counterparts (Figs.~\ref{fig:croissant}c and d). This is partly explained by thickness maps that slightly differ from the initial design (see Supplementary Fig.~S1).
It also appears that the incident wave-fields are not perfectly linear. The glue layer below the piezoelectric strip is not perfectly homogeneous. Nevertheless, the distortion of the incident wave-field can be taken into account in the computation of the Strehl ratio. Indeed a complementary medium shall, in theory, reproduce exactly at its output the wave-field at its input. Hence, by subtracting the incident phase to the recorded wave-field, the Strehl parameter can be made independent of the incident wave-field:
\begin{equation}
\label{eq:Strehlratio2}
    \mathcal{S}(x) = \left |< e^{i [\phi(x,y)-\phi(x=0,y)]} >_y\right |,
\end{equation}
The result is displayed in Fig.~\ref{fig:croissant}h. The experimental Strehl ratios exhibit an evolution similar to the numerical predictions (Fig.~\ref{fig:croissant}e). However, in the complementary plate, it does not reach the expected value of 1 but rather saturates around $\mathcal{S}\sim0.7$. The cloaking effect is thus less spectacular in the experiment than in the numerical simulation. Several experimental limitations can explain this difference (see Supplementary Fig.~S1). 
\al{First,} the thickness of the negative index areas is not perfectly homogeneous. Second, the engraved anti-object and object are not exactly mirror from each other. Hence, the wave-front recombination at the device output cannot be optimal. These devices suffer from the same limitations than previous cloaking realisations reported in the literature~\cite{stenger2012experiments,jin2016gradient,alu2005achieving,schurig2006metamaterial}: The cloaking performance is hampered by manufacturing imperfections.}

\alex{Nevertheless, the results shown} here constitute a first experimental proof of concept of the scattering cancellation through complementary media for elastic waves \alex{and even, to our knowledge, in wave physics}. \alex{Furthermore, the use of natural backward Lamb waves shows the advantage of being free of the meticulous conception of a resonant metamaterial. This strategy has allowed us to overcome the energy dissipation issues} generally encountered in such man-made media. \ree{As rigorously demonstrated by Nguyen~\cite{Nguyen2017}, cloaking by an anti-object is not restricted to a plane wave illumination and/or a finite size scatterer. Supplementary Fig.~S2 illustrates} the generality of this concept by considering the case of complementary bands~\cite{pendrycomp} insonified with a point-like source. 
Each band is shown to annihilate wave diffraction from the other. The overall effect is as if a section of space was removed from the experiment. \ree{Note that this striking feature has also been predicted in three-dimensional geometries (\textit{e.g} annular lens~\cite{pendry2003focusing}) but it requires not only a complementary but also a heterogeneous refractive index distribution~\cite{Nguyen2017}.}

Another interesting ability of cloaking is also to isolate an object from the external \al{environment} with no possible interaction. In other words, no wave can access the object from the outside, and no wave generated by the object can get out of the cloak. In the next section, we show that this is partially the case in complementary media. \alex{A part of the wave generated by a source inside a complementary medium remains trapped inside. To highlight this striking phenomenon, Lamb wave propagation is investigated around a \al{double} perfect corner (Fig.~\ref{fig:SchemaCroissantAndPC}b).}

\vspace{5 mm}

\begin{figure*}[tb]
  \centering
  \includegraphics[width=\textwidth]{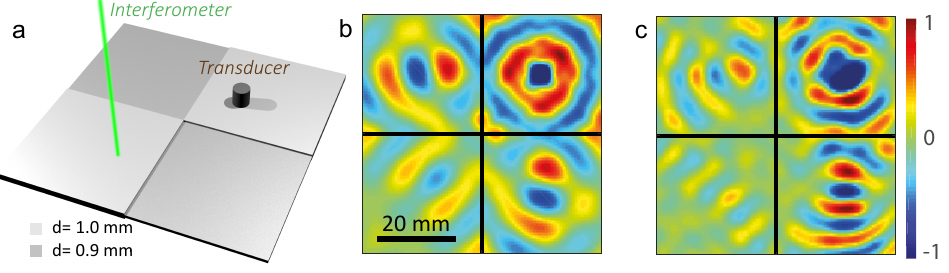}
  \caption{\label{fig:PC} \alex{\textbf{Wave focusing in a perfect corner.}} \textbf{a} Experimental configuration. \textbf{b} Real part of the normal displacement wave field simulated at the surface of the plate at the crossing frequency $f_c=3.32$ MHz \re{(SimSonic3D software~\cite{bossy2004three}, www.simsonic.fr)}. \textbf{c} Real part of the normal displacement wave field experimentally measured at the surface of the plate at the crossing frequency $f_c=3.46$ MHz.}
\end{figure*}

\noindent \textbf{\alex{Trapping by a double perfect corner.}}
The double perfect corner device is composed of four quadrants of opposite refractive index (Fig.~\ref{fig:PC}a). A point source at point A emits a \fab{diverging wave} in one of the positive index quadrant. This wave-field is then negatively refracted at both interfaces in the negative index quadrants. Thus the wave, by successive negative refraction phenomena, travels around the double corner creating images of the source in each quadrant, named A, B, C and D in Fig.~\ref{fig:SchemaCroissantAndPC}c.
This trapping phenomenon has been highlighted, both theoretically and numerically, for electromagnetic waves~\cite{guenneau2005perfect} and for flexural waves~\cite{guenneau2010perfect}. 

A numerical simulation of the \al{double} perfect corner for Lamb waves is first performed~\cite{bossy2004three,simsonic} (see Methods). \al{As before, a duralumin plate is considered, with positive and negative index parts of thickness 1 mm and 0.9 mm respectively.} A normal displacement point source is placed on the top right quadrant at $15$ mm from both interfaces. It generates a cylindrical forward $S_2$ Lamb wave.  At each interface, a conversion into the backward mode $S_{2b}$ is expected. 
Figure~\ref{fig:PC}b shows \al{the simulated displacement field} at the crossing frequency. The trapping of Lamb modes around the corner is illustrated by a refocusing of waves in each quadrant. Admittedly, the focusing in the third quadrant is not as well resolved as in the other quadrant. \al{This unwanted effect is inherent to the device as explained by a ray approach depicted  in Fig.~\ref{fig:SchemaCroissantAndPC}b:} (i) \al{Part} of the emitted rays straightly propagate towards the edges of the plate and get absorbed by perfectly matched layers (red arrows); (ii) \al{Part} of the negatively refracted rays are also lost in the second and third quadrants (\corr{orange arrows}). Therefore, only half of the angular spectrum of the incident wave-field (blue arrows) contribute to the refocusing in the the opposite quadrant.

Following this numerical study, a double perfect corner is manufactured on a duralumin plate \al{by chemical etching} (see Methods). The erosion process being imperfect, a control of the thickness is \alex{necessary} and the corresponding thickness map is shown in Supplementary Fig.~S1.
The source is a piezoelectric transducer glued on the top right positive index quadrant at $15$ mm from both interfaces (see Methods). The normal displacement is measured at the plate surface using a \alex{photorefractive} interferometer. \al{The wave-field filtered at the crossing frequency is shown} in Fig.~\ref{fig:PC}c.
\al{As predicted, refocusing} occurs in each quadrant. The resulting wave field is closely similar to the numerical result displayed in Fig.~\ref{fig:PC}b obtained with perfectly matched layers. The reflections on the plate edges are therefore insignificant in the experiment because of the overall dimension of the plate (see Methods). 
The residual discrepancy between the numerical and experimental wave-fields is explained by: (\textit{i}) a different thickness ratio ($d_2/d_1=0.86$) that implies a shift in terms crossing frequency ($f_c=3.46$ MHz instead of 3.32 MHz) and \corr{corresponding wavelength ($\lambda_c=\al{10}$ mm instead of 12.2 mm)};  (\textit{ii}) thickness fluctuations in the eroded areas (see Supplementary Fig.~S1).
\begin{figure*}
  \centering
  \includegraphics[width=\textwidth]{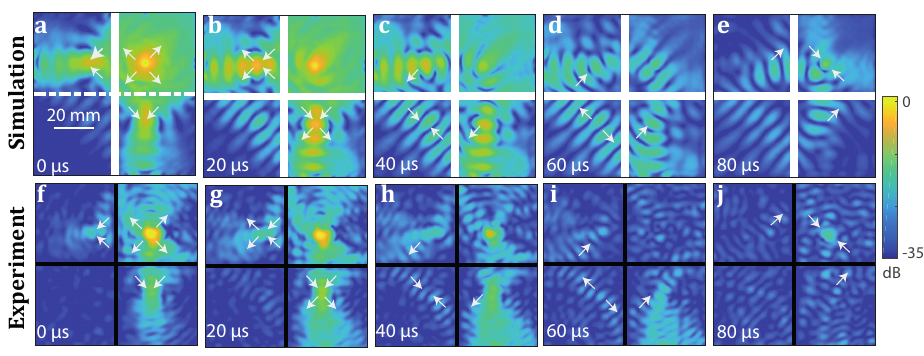}
  \caption{\alex{\textbf{Trapping a wave-packet around a perfect corner}. Snapshots of a wave packet \rev{($\Delta f \sim 0.02$ MHz)} propagating around the double perfect corner. \ree{\textbf{a}-\textbf{e}} Numerical result for $f_c=3.32$ MHz \re{(SimSonic3D software~\cite{bossy2004three}, www.simsonic.fr)}. \ree{\textbf{f}-\textbf{j}} Experimental result for $f_c=3.46$ MHz. \rev{The color scale is in dB. White arrows indicate in each panel the trajectory followed by the wave-packet.}}}\label{fig:PC_temp}
\end{figure*}

While a study of the wave-field in the frequency domain nicely shows the refocusing of waves in each quadrant, \al{it is interesting to observe the field in the time domain in order to assess the \rev{circulation}} of Lamb waves around the double perfect corner. To that aim, an inverse Fourier transform of the wave-field is performed over a frequency bandwidth of \rev{$0.02$ MHz} around the crossing frequency. The corresponding temporal wave-field is shown \rev{at different times of flight} in Fig.~\ref{fig:PC_temp}a-e. The images of the source located in quadrant A are observed in quadrants B and C at time $t=20~\mu$s (Fig.~\ref{fig:PC_temp}b), before recombining in quadrant D at $t=40$ $\mu$s (Fig.~\ref{fig:PC_temp}c). As expected, the negatively refracted waves then go through quadrants B and C [$t=60~\mu$s, Fig.~\ref{fig:PC_temp}d] before refocusing at the initial source location in quadrant A at time $t=80~\mu$s (Fig.~\ref{fig:PC_temp}e). \rev{The amplitude loss ($\sim$ -10 dB) exhibited by the refocused wave compared to the source is explained by the fact that only a part of the emitted rays can be trapped by the double negative corner (blue arrows in Fig.~\ref{fig:SchemaCroissantAndPC}b). }

The temporal behavior of the wave-field around the double perfect corner is now investigated experimentally \rev{(Fig.~\ref{fig:PC_temp}f-h). A qualitative agreement is obtained with the numerical prediction (Fig.~\ref{fig:PC_temp}a-e).} The emitted wave [$t=0$ $\mu$s, Fig.~\ref{fig:PC_temp}f] refocuses in quadrants B and C [$t=20$ $\mu$s, Fig.~\ref{fig:PC_temp}g]. However, the displacement \alex{amplitude} is lower in quadrant C than in quadrant B. \alex{This is explained by the fact that the thicknesses of quadrant B and C slightly differ (see Supplementary Fig.~S1), \al{resulting in} different transmission coefficient\al{s} at the interface. Moreover, some step irregularities probably induce spurious reflections at the interface between quadrants A and C. Nevertheless, negatively refracted waves recombine in quadrant D ($t=40$ $\mu$s, Fig.~\ref{fig:PC_temp}h). The wave packets then go back to quadrants B and C ($t=60$ $\mu$s, Fig.~\ref{fig:PC_temp}i) \rev{before refocusing at the initial source location in quadrant A for the return trip time $t=80$ $\mu$s (see Fig.~\ref{fig:PC_temp}j). Strikingly, the refocusing process is even better at the initial source location than in the other quadrants. A potential explanation is the spatial reciprocity between the clock-wise and anti-clockwise paths that is robust to plate imperfections and makes them interfere constructively at the initial source location \re{(see Supplementary Fig.~S3)}. This phenomenon is reminiscent of the coherent back-scattering phenomenon that arise in cavities~\cite{Rosny2000,Catheline} or random media~\cite{Wolf1985,Albada1985}. This experiment thus demonstrates the ability of trapping a wave-packet around a double negative corner.}}

{Lamb wave focusing by negative refraction, as observed for the flat lens~\cite{legrand2018negative} or for the perfect corner, is diffraction limited: \rev{The transverse dimension of each focal spot is of the order of a half-wavelength.}
{Unlike Pendry's initial proposal~\cite{pendry2000negative}, the Lamb mode negative refraction \rev{process} \rev{is actually only efficient for the propagating component of the wave-field}.}} \ree{Indeed, the anomalous resonance mechanism needed to replicate the evanescent field of the source in each quadrant requires more stringent conditions than the mere anti-symmetric phase velocity distribution considered in this paper~\cite{Milton2005,McPhedran2020,Deng2020}. In the next section, a strategy is proposed to circumvent this issue by designing an elastic superlens that combines the negative refraction phenomenon~ and the metalens concept~\cite{lemoult2010resonant}}.

\vspace{5 mm}

\noindent \textbf{\alex{Superlensing by combining negative refraction and metalens concepts.}}
Lemoult \textit{et al.} introduced the concept of metalens~\cite{lemoult2010resonant}, an arrangement of sub-wavelength resonators that allow to convert the evanescent components of a source into propagating waves ~\cite{lemoult2010resonant,lemoult2011acoustic}. Initially coupled to a time-reversal mirror in order to produce sub-wavelength focusing at the source location, our idea is here to couple the metalens concept to the negative refraction phenomenon. Time reversal and negative refraction are actually intimately linked processes~\cite{pendry2008time}. Nevertheless, \rev{a first} difference is that the object and imaging planes are distinct in a negative refraction scheme\rev{. Unlike time-reversal that implies back-focusing on the source, the evanescent components of the object wave-field should be here recombined in an} imaging plane~\cite{krishnan2001evanescently}. \rev{The second difference lies in the nature of the metalens. While time reversal experiments rely on a resonant metalens in order to encode the spatial details of the source in time~\cite{lemoult2010resonant,lemoult2011acoustic}, the negative refraction superlens here only requires an efficient converter of the evanescent components of the source into propagating waves.} 

In this paper, the following strategy is thus adopted: \al{Introducing} two metalenses in the vicinity of the object and imaging planes (see Fig.~ \ref{fig:SchemaCroissantAndPC}c). The first metalens will convert the evanescent components of the object into propagating waves. \alex{The corresponding} wave-field is reproduced in an imaging plane thanks to a negatively refracting slab. By virtue of spatial reciprocity theorem, \rev{the second metalens in the imaging plane should be made central symmetric to the first one with respect to the center of the device. Only on this condition will the second metalens back-convert the negatively refracted waves} into the evanescent components of the object in the imaging plane. 

\rev{The negative refracting lens considered here is \al{of 0.92 mm thickness and  50 mm width} (Fig.~\ref{fig:SchemaCroissantAndPC}c). The object and imaging planes are at 25 mm from the negative refracting slab interfaces. A point source is placed in the vicinity of the object plane ($x=20$ mm).} The first meta-lens should be designed for the incident forward $S_2$ Lamb mode. As demonstrated in previous works~\cite{xeridat2011etude,dubois2014controle,Dubois2017}, an efficient sub-wavelength \rev{scatterer} for Lamb waves is the resonant blind hole. \al{The perforation must be symmetric in order to avoid any conversion into anti-symmetric modes. The holes depth are chosen to maximize the reflection of $S_2$ mode in itself. An optimal thickness ratio of 0.76 between the plate and the holes is found using the semi-analytical model described in Ref.~\onlinecite{legrand2018negative} (see Supplementary Fig. \re{S4}). The holes diameters are set to 2 mm in order to meet the following criteria: (\textit{i}) being sufficiently small compared to the $S_2$ mode wavelength ($\lambda_c=12.5$ mm) \al{so that a large contribution of evanescent components of the source is converted into propagating waves}; (\textit{ii}) being large enough to maintain a sufficiently high scattering cross-section. \rev{While periodicity is not a prerequisite for the metalens, a periodic arrangement of blind holes is nevertheless considered here for sake of simplicity. The inter-hole distance \rev{is} fixed to $1$ mm in order to \rev{get an efficient conversion} between the evanescent and propagating components of the $S_2$ mode over the transverse spatial frequency range $k_c<|k_y|<k_{S1}$ (see Fig.~\ref{fig:CdispDural}). The second meta-lens, placed in the imaging plane, is made identical to the first one to meet the aforementioned reciprocity requirement between the two meta-lenses.} The whole system is simulated by means of a FDTD code~\cite{bossy2004three,simsonic} (see Methods).} 

\begin{figure*}[t]
  \centering
  \includegraphics[width=\textwidth]{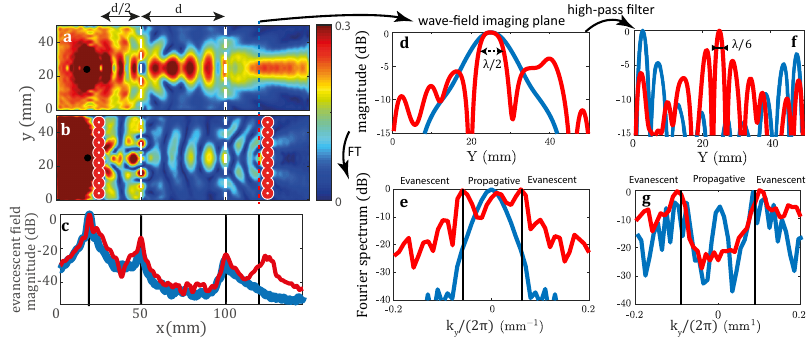}
  \caption{\alex{\textbf{Transporting an evanescent wave-field in the far-field by means of an elastic superlens.}} \textbf{a}, \textbf{b} Normal displacement magnitude at plate surface without (\textbf{a}) and \alex{with (\textbf{b}) metalenses} \re{(SimSonic3D software~\cite{bossy2004three}, www.simsonic.fr)}. \textbf{c} \rev{Evolution along the $x$-direction of the evanescent component magnitude of the wave-field \rev{(averaged over $|k_y|>k_c$)} without (blue) and with (red) metalenses.} \alex{\textbf{d} Normal displacement magnitude along the $y$-axis in the imaging plane ($x=120$ mm) (blue) without and (red) with metalenses. \textbf{e} Spatial Fourier transform of the image wave-field without (blue) and with (red) metalenses. \textbf{f} High-pass filter applied to the image wave-field in order to point out the evanescent component contribution to the focusing. \textbf{g} Spatial Fourier transform of the filtered image wave-field. In panels \textbf{d}-\textbf{g}, the displacement magnitude is shown in dB.}}\label{fig:superlens}
\end{figure*}

\ree{For sake of comparison, a reference numerical simulation of the same system without the metalenses is performed.} Figures \ref{fig:superlens}a and b display the modulus of the normal displacement \rev{$u(x,y)$} \al{for the reference and metalenses devices, respectively,} at the crossing frequency. 
\al{As already investigated experimentally~\cite{philippe2015focusing,legrand2018negative}, the negative refracting slab in Fig.~\ref{fig:superlens}a shows a double focusing process in the slab and after it.} The spatial profile of the focal spot in the imaging plane is displayed in Fig.~\ref{fig:superlens}d. Its full width at half maximum (FWHM) $\delta$ is of 11.2 mm ($\al{\delta} \sim \lambda_c$). This spatial extension is limited by the numerical aperture of the negative refracting lens: $\delta \sim \lambda_c/ (2 \sin \theta)$ where $\theta \sim 40^{\textrm{o}}$ is the maximum angle under which the lens aperture is seen from the point source. 
In presence of metalenses, a double refocusing process is also observed in Fig.~\ref{fig:superlens}b. As before, the focal spot inside the negative refracting lens seems diffraction-limited. \re{This observation indicates that, with or without metalens, the negative refraction slab mostly supports propagating waves.} However, the focal spot in the imaging plane looks much thinner in presence of the second metalens, compared \al{to} the reference wave-field in Fig.~\ref{fig:superlens}a.  \fab{This \re{striking feature} is confirmed by Fig.~\ref{fig:superlens}b that displays} the displacement magnitude of this focal spot across the imaging plane. Its FWHM $\delta$ is close to $\lambda_c$ which implies a numerical aperture $\sin \theta \sim 1$. 

\rev{To assess the relative weight of propagating and evanescent components in the imaging plane, the Fourier transform of the wave-field $u(x,y)$ can be performed along the $y$-direction: \rev{$\hat{u}(x,k_y)=\int u(x,y) e^{-i k_y y} dy$}~\cite{Goodman}. The propagating components of the wave-field are associated with a wave vector $\mathbf{k}=(k_x,k_y)$ whose transverse component $k_y$ is smaller than the wavenumber $k_c=2\pi/\lambda_c$ of the $S_2$ Lamb mode and whose longitudinal component $k_x=\sqrt{k_c^2-k_y^2}$ is real. The evanescent part of the wave-field corresponds to transverse components $k_y$ larger than $k_c$ and a purely imaginary longitudinal component $k_x=i\sqrt{k_y^2-k_c^2}$. Figure~\ref{fig:superlens}e shows the modulus of $\hat{u}(x,k_y)$ in the imaging plane. In absence of metalens, the Fourier spectrum of the wave-field is bounded between $-\al{k_c}\sin(\theta)$ and $k_c \sin(\theta)$. It confirms the limited collection angle of the original negative refraction lens and the absence of evanescent components in the imaging plane in that case. On the contrary,} in presence of metalenses, both propagating ($|k_y|<\al{k_c}$) and evanescent components ($|k_y|>\al{k_c}$) contribute to the wave-field in the imaging plane. 

\rev{The enhancement of the evanescent components in the imaging plane} can be accounted for by the \al{effect} of the second metalens. Not surprisingly, an array of sub-wavelength \rev{scatterers can induce sub-wavelength details} in the near-field. The question is to determine \rev{whether} this evanescent wave-field contributes to the focusing process in the imaging plane. To do so, a spatial high-pass band filter has been applied to get rid off the propagating part of the wave-field and only focus on its evanescent part [see Fig.~\ref{fig:superlens}g]. The corresponding focal spot is obtained by an inverse Fourier transform and its magnitude is displayed in Fig.~\ref{fig:superlens}f. \alex{\al{In presence of metalenses,} a sharp focal spot is obtained exactly at the location of the source image with a FWHM $\delta \sim \lambda_c/6$. The strong secondary lobes at -3 dB are induced by the cut-off frequency of the high pass filter. Nevertheless, this result shows that the negative refraction and metalens phenomena are perfectly complementary. \al{Indeed, they provide: (\textit{i}) a transport of the evanescent wave-field over a distance much larger than the wavelength ( $\sim 8\lambda_c$ in the present case);  (\textit{ii}) a proper recombination of the evanescent components in the imaging plane, thereby leading to sub-wavelength focusing.} This numerical proof-of-concept \al{thus} paves the way towards a future experimental implementation of an elastic superlens.}

\rev{One perfectible point remains, however, the loss of the evanescent components ($\sim 12$ dB) compared to the propagating wave-field highlighted by Fig.~\ref{fig:superlens}e. This loss is not due to absorption or impedance mismatch at the interfaces of the negative refraction lens. It can be accounted for by the imperfect conversion ($\sim$0.5 in amplitude) between the evanescent components into propagating waves operated by each metalens in the source and imaging planes. The resolving power of the superlens obviously depends on the quality of this conversion. Further work is thus needed to optimize it through a careful design of the metalens.}

\vspace{5 mm}

\noindent {\large \textbf{Discussion}} \\

In this paper, \al{backward Lamb modes have} been taken advantage of to design and implement devices based on the physics of negative refraction and complementary media. A previous analytical study and numerical simulations have allowed the optimization of such devices and demonstrated the merits of this strategy. In particular, a material like duralumin displays an excellent conversion between forward and backward Lamb modes at a thickness step. \rev{Moreover, these devices do not require any periodicity and do not rely on any resonant phenomena, which make them particularly robust to absorption losses in contrast with meta-materials.} 

These performances have been qualitatively confirmed by experimental realizations of such devices by means of laser interferometry. Nevertheless, these experiments have also pointed out the extreme sensitivity of complementary media \al{to} manufacturing imperfections. The etching methods used in this paper, namely chemical etching and die sinking electrical discharge machining, have shown some limits, in particular for tailoring sharp corners and guaranteeing a perfectly homogeneous plate thickness over a few decimeters. To cope with these issues, additive manufacturing, in particular selective laser melting \al{(or laser powder bed fusion)}, seems to be particularly promising. Other micro-machining methods used for semiconductors or MEMS, such as laser-assisted wet etching~\cite{Wang2020}, could provide a sharper design but they would require to work on smaller plate dimensions, hence higher frequencies and larger dissipation losses. In that perspective, alternative materials more adapted to each manufacturing technique could be used such as silicon~\cite{Prada2009} or aluminium~\cite{Gruensteidl2018}. Note, however, that this change of material can be detrimental to the conversion efficiency between forward and backward Lamb modes at each thickness step~\cite{legrand2018negative}.  
\alex{On a more fundamental side, the striking properties of complementary media can be leveraged by means of transformation optics~\cite{leonhardt2006optical,Pendry2006,Pendry2012}. Indeed, the ratio of plate thickness $d$ to wavelength $\lambda$ determines the effective stiffness of the plate and the phase velocity of the mode. As shown by Lefebvre et al.~\cite{lefebvre2015experiments} for flexural waves, one can tune the local phase velocity with the thickness of the plate. \al{Another option consists in using surface bonded slice lenses~\cite{Tian2017}.} Graded index devices can thus be designed by means of conformal transformations~\cite{Tang2020,Zhao2020}. Applied to any complementary medium, it can lead to the design of a whole set of cloaking devices~\cite{lai2009complementary,Guenneau2021}. }  

\alex{With regards to the superlens, one open question remains the existence of resonant sub-wavelength Lamb modes at a thickness step. Analogous to the role of surface plasmons in the perfect lens of Pendry~\cite{pendry2000negative}, they would allow the transport of the evanescent components across a negative refracting lens made of two thickness steps. \ree{These surface resonant states~\cite{pendry2000negative,pendry2003focusing,Li2004,Ambati2007} and the associated anomalous resonance~\cite{Milton2005,Nguyen2015} do exist for both acoustic and electromagnetic waves.} \ree{However, their existence has not yet been proven experimentally for elastic waves, albeit theoretically predicted~\cite{McPhedran2020,Deng2020}.} For the more specific case of Lamb waves, such resonant states seem to arise at the thickness steps in the elastic superlens. \al{An enhancement of the evanescent components of the wave-field is actually observed at each thickness step of the negative refracting lens (see Fig.~\ref{fig:superlens}c).} Further investigations are thus needed to understand the nature and role of these resonant states at the interface of the elastic superlens. On the theoretical side, the existing analytical model~\cite{legrand2018negative} should be extended to the case of evanescent incident wave-fields. On the numerical side, finite element modeling would be more adapted than FDTD simulations because of the resonant feature of a perfect lens. \al{An adaptive meshgrid would also enable a finer spatial sampling in the vicinity of thickness discontinuities.} On the experimental side, a monochromatic and spatially-selective generation of Lamb waves~\cite{Gruensteidl2018} would be required to excite selectively such resonant modes.} 

\alex{In this work, we have highlighted both numerically and experimentally the elegant physics of complementary media in the context of guided elastic waves. Lamb waves are actually perfect candidates to observe negative refraction phenomena since, under certain conditions, thickness steps on a plate can generate a very efficient conversion between forward and backward Lamb modes. By designing complementary media as an arrangement of thickness steps over a plate, the ability of cloaking a part of the space or of trapping waves around some singular points has been demonstrated. Despite manufacturing imperfections, this is a first experimental proof-of-concept of complementary media for elastic waves. At last, a numerical study has paved the way towards a new route for the design of an elastic superlens. By combining negative refraction with the metalens concept, the evanescent field of a source can be transported in the far-field and recombined in an imaging plane by means of a negative refracting lens. However, two main issues remain to be solved before implementing experimentally such \al{an} elastic superlens: (\textit{i}) A more elaborated manufacturing strategy for sharply tailoring the metalens and the thickness steps ; (\textit{ii}) A better theoretical understanding and experimental harnessing of resonant surface modes that have been highlighted at each thickness step. From a more applied point-of-view, Lamb waves draw nowadays increasing attention for the design of new electro-acoustic devices in electrical engineering or acoustic sensors in MEMS technology. They are also commonly used for non-destructive testing in the aviation, automobile or nuclear power industries. For all these applications, the control of Lamb waves is essential. In that perspective, the ability to cancel their propagation in some parts of the space or to focus them at a deep sub-wavelength scale is of a great interest whether it be for imaging, sensing or filtering applications.}

\vspace{5mm}

\noindent {\large \textbf{Methods}} \\
{\small
\noindent \textbf{\alex{Numerical simulations.}} \alex{Numerical simulations have been performed with the \alex{FDTD} Simsonic software~\cite{bossy2004three} to study the propagation of elastic waves across a duralumin plate of density $\rho=2795$ kg.m$^{-3}$, longitudinal \alex{wave velocity} $c_L = 6400$ m.s$^{-1}$ and \alex{transverse wave velocity} $c_T = 3120$ m.s$^{-1}$.
The plate dimensions and mesh size are given in Tab.~\ref{table}. Perfectly matched layers (PML)s are applied at the edges of the plate in order to avoid spurious reflections. 
{\small
\begin{table}[h!]
\center
\begin{tabular}{c|c|c}
Simulated   & Complementary  & Superlens  \\
device & media & \\
\hline
\hline
Dimension [mm$^2$]& $200\times200$  &  $ 150 \times 150 $ \\
Thickness [mm] & 1 - 0.9 & 1 - 0.9 \\
Mesh size [mm] & 0.05 & 0.04 \\
\hline
\end{tabular}
 \caption{\label{table}\textbf{Simulation Parameters}.}
\end{table}
}} 

\alex{Depending on the simulated device, the source is either point-like (\al{double} perfect corner, superlens) or linear (anti-object). It consists in a normal displacement source applied at the surface of the plate. The excitation function is a pulse of 3.3 MHz central frequency with -6 dB bandwidth of 100\%. 
The normal displacement of the induced wave-field is then recorded over $110~\mu$s at the surface of the plate with a spatial period $\delta = 0.2~ \text{mm}$. }\\

\noindent \textbf{\alex{Experiments.}}
\alex{For each experiment, a $200\times200$ mm$^2$ duralumin plate is used. Such large dimensions \al{minimize} reflections on the edge of the plate during the recording time. The plate, which initially displays an homogeneous thickness of 1 mm, is engraved in the negative index parts to obtain a thickness of $0.9$ mm. The scattering object in the cloaking device is engraved by die sinking electrical discharge machining. The other negative index area (the band including the anti-object and the double corner) are engraved by chemical etching using iron per-chloride. The actual thickness maps for each device are displayed in Supplementary Figs.~S1 and S2.}

\alex{In the cloaking experiment, the source is a $10$ mm wide, $100$ mm long and $1$ mm thick piezoelectric strip. In the double perfect corner experiment, the source is a 7 mm diameter piezoelectric transducer (Olympus V183-RM). In each experiment, the source is glued on to the thick part of the plate using Phenyl salicylate. The excitation signal is a $5$-$\mu$s-long linear chirp sweeping a frequency spectrum ranging from $3.1$ to $3.6$ MHz.}
 
\alex{The normal displacement of the wave field is measured at the surface of the plate using a homodyne interferometer with a photo-refractive crystal (Sound$\&$Bright, TEMPO 1D) for the reference plate and a homodyne interferometer (Sound$\&$Bright, Quartet) for the plate with the anti-object because of the strong light scattering induced by the etched area. }\\

\noindent \textbf{\alex{Data post-processing.}}
\alex{A temporal Fourier transform is then applied to the recorded wave-fields in order to study each device at the crossing frequency $f_c$ between the forward and backward modes: $f_c = 3.32$ MHz (in numerical simulations), $f_c=3.37$ MHz for the cloaking reference plate, $f_c=3.29$ MHz for the anti-object cloaking plate, and $f_c=3.46$ MHz for the perfect corner plate. This wave field is then filtered spatially using a low pass filter in order to isolate the contribution of the $S_2$ and $S_{2b}$ modes. The spatial frequency cutoff of this filter is $0.25$ mm$^{-1}$.}

\alex{In the superlens simulation, the propagating components of the wave-field are filtered by means of an additive high-pass filter whose spatial frequency cutoff is 0.085 mm$^{-1}$ [see Fig.~\ref{fig:superlens}(e)]. The attenuation applied to the propagating component of the wave-field is of $20$ dB.}}

\clearpage 

\clearpage

\renewcommand{\thetable}{S\arabic{table}}
\renewcommand{\thefigure}{S\arabic{figure}}
\renewcommand{\theequation}{S\arabic{equation}}
\renewcommand{\thesection}{S\arabic{section}}

\setcounter{equation}{0}
\setcounter{figure}{0}

\begin{center}
\Large{\bf{Supplementary Information}}
\end{center}

This document provides further information on: (\textit{i}) the mapping of the plate thickness in each experimental device; (\textit{ii}) the study of complementary bands; (\textit{iii}) the design of blind holes in the numerical study of the superlens.

\section*{{Plates thickness mapping}}

In order to evaluate the quality of the manufactured plates, a local measurement of the plate thickness should be performed. To that aim, a \al{map} of the ZGV resonance frequency (Fig.~2) is achieved by means of a heterodyne interferometer and a pulsed Nd:Yag laser whose wavelength is 1064 nm (Centurion, Quantel)~\cite{prada2005laser}. 

For the cloaking plates, the ZGV frequency measured on the thick part of the reference plate ($f^{+}_{zgv}=2.888$ MHz) is higher than the one measured on the complementary plate ($f^{+}_{zgv}=2.812$ MHz). The reference plate is thus thinner by a factor $0.97$. This difference implies a slightly different wave-length $\lambda_c$ exhibited by the $S_2$ and $S_{2b}$ modes at the crossing frequency in each plate (see Fig.~3 of the accompanying paper).

\begin{figure*}[ht]
  \centering
  \includegraphics[width=17cm]{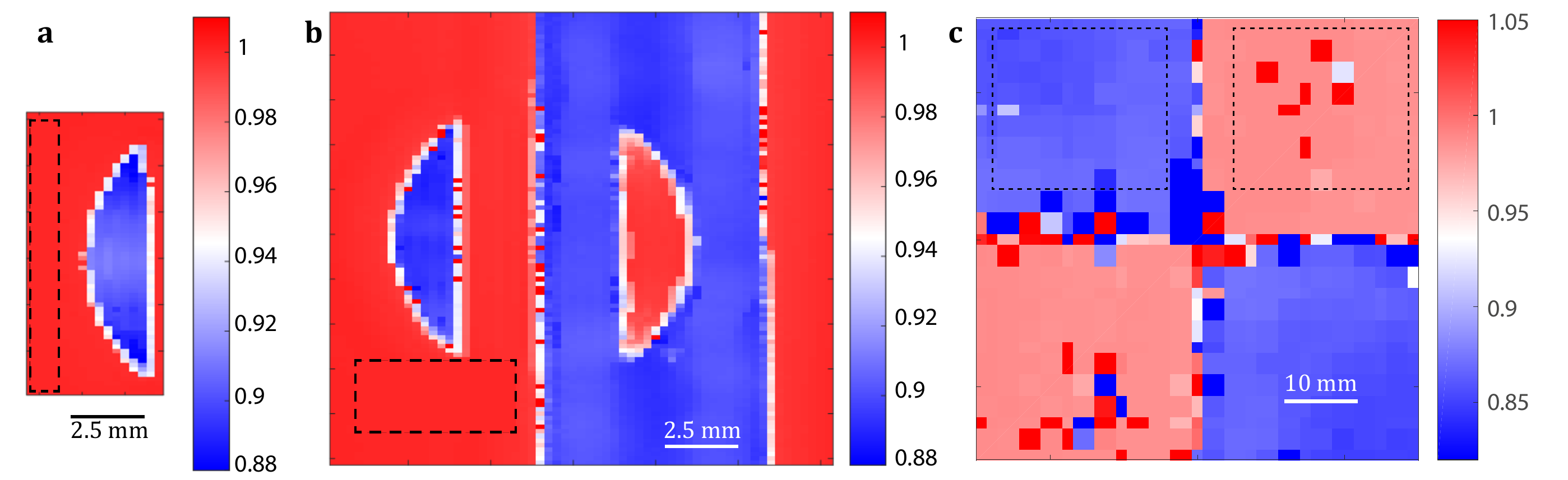}
  \caption{\textbf{Maps of the plates thickness.} \textbf{a} Reference plate. \textbf{b} Complementary plate. \textbf{c} Double perfect corner plate. In each case, the measured thickness is normalized by its mean value ($d_1=$1 mm) in the thick part.}\label{fig:suppl:cloakMaps}
\end{figure*}

Figures~\ref{fig:suppl:cloakMaps}a and b display the ZGV frequency across the reference and complementary plates normalized by their respective $f^{+}_{zgv}$. This quantity thus yields the thickness across the plate normalized by its value \al{obtained from an average over the black dotted rectangle}. In each case, the thickness ratio between the thin and thick parts of the plate ranges between 0.88 and 0.91. Note that, in the complementary plate, the object and anti-object are not strictly mirror from each other. This partly explains the lower cloaking performance obtained experimentally compared to the numerical prediction (see Fig.~3 of the accompanying paper).

A relative thickness map of the perfect corner device has been measured with the same technique. The result is displayed in Fig.~\ref{fig:suppl:cloakMaps}c. The mean thickness ratio is $d_2/d_1=0.86$. This value implies a crossing frequency $f_c=3.46$ MHz between the forward $S_2$ mode in the thick part and the backward $S_{2b}$ mode in the thin part of the plate.

\section*{{Complementary bands}}

\begin{figure*}[ht]
  \centering
  \includegraphics[width=0.8\textwidth]{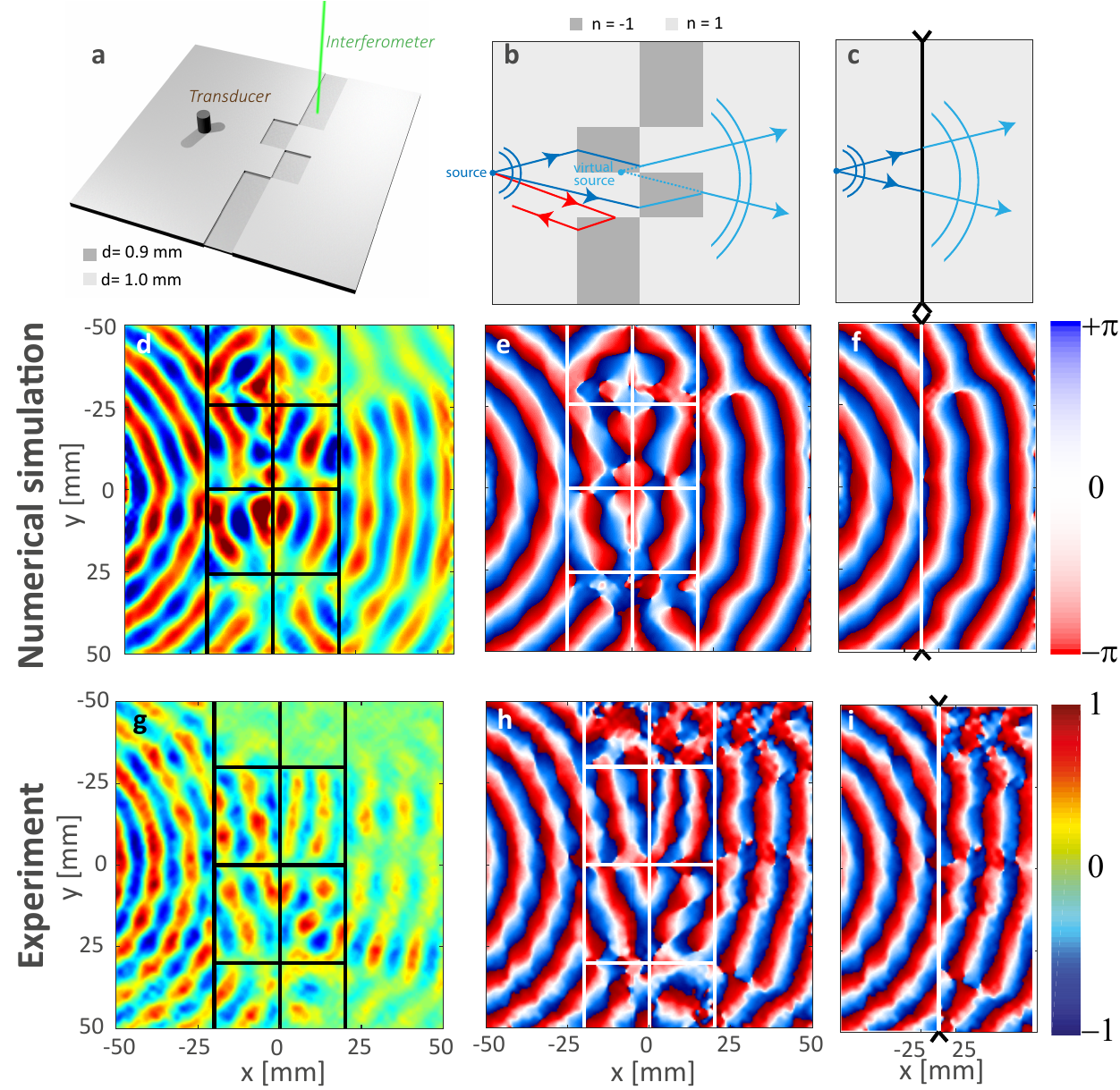}
  \caption{\textbf{Cancelling wave diffraction over a region of space by means of complementary bands.} \textbf{a} Experimental configuration.  \textbf{b} An alternative pair of complementary bands annihilates the effect of the other. Wave does not follow a straight line path in each layer, but the overall effect is as if a section of space thickness $2L$ were removed from the experiment. \textbf{c} The latter effect is highlighted by juxtaposing the wave-fronts upstream and downstream to the complementary bands. \textbf{d}-\textbf{f} Numerical simulation of the complementary bands: Real part (\textbf{d}) and phase (\textbf{e}) of the normal displacement wave-field at the crossing frequency $f_c = 3.32$ MHz.  (\textbf{f}) Juxtaposition of the wave-fronts at the input and output of the complementary bands as in \textbf{c}. \textbf{d}-\textbf{f} Experimental implementation of the complementary bands: Real part (\textbf{d}) and phase (\textbf{e}) of the normal displacement wave-field recorded at the crossing frequency $f_c=3.37$ MHz. \textbf{f} Juxtaposition of the wave-fronts at the input and output of the complementary bands as in \textbf{c}. }\label{fig:comp}
\end{figure*}

A negative refracting slab can be as a piece of negative space: It annihilates the diffraction of waves propagating over an equal thickness of opposite index. This striking property gave birth to the concept of complementary media. Two slabs of material of equal thickness and placed adjacent to one another annihilate each other if one is the negative mirror image of the other, the mirror being taken to lie on the interface between the two slabs. 

The first configuration proposed by Pendry~\cite{pendrycomp} consisted in associating two complementary bands of thickness $L$, an example of which is given by Fig.~\ref{fig:comp}b. The phase accumulated by the wave during its travel through the first band is exactly compensated by the complementary band. At the output of this system, the transmitted wave-front is perfectly analogous to the incident one. The overall effect is as if a layer of space of thickness 2$L$ had been removed from the experiment (Fig.~\ref{fig:comp}c). This property holds whatever the incident wave-field. For the particular case of a point-like source, the complementary bands translate this source into a virtual one shifted by a distance $2L$ (Fig.~\ref{fig:comp}b).

In this work, we have implemented this idea for guided elastic waves. The system studied here is a 1 mm\alex{-}thick duralumin plate excited by a piezoelectric transducer that emits the forward $S_2$ mode as a cylindrical wave (Fig.~\ref{fig:comp}a). The designed complementary bands consists in an arrangement of thickness steps over the plate. The positive and negative index areas corresponds to a plate thickness of 1 and 0.9 mm, respectively. At the crossing frequency and at each thickness step, there is a conversion between the forward $S_2$ and backward $S_{2b}$ modes (see Fig.~2). The overall thickness $2L$ of the complementary bands is of 40 mm.

These devices are first investigated numerically using a FDTD code~\cite{bossy2004three,simsonic}. The simulation parameters are described in the Methods section of the accompanying paper. A normal displacement pulse is applied to the point source placed at a distance of 25 mm from the complementary bands. The normal displacement induced at the plate surface is recorded over a time length $\Delta t=110$ $\mu$s. A spatio-temporal filter described in the Methods section is then applied to the recorded wave-field in order to isolate the $S_2$ and $S_{2b}$ mode contributions. Figure~\ref{fig:comp}d and e show the corresponding wave field and its phase at the crossing frequency $f_c=3.32$ MHz. The incident wave is negatively refracted in the thinnest parts of the first complementary band, resulting in a strongly distorted wave-front at its output ($x=0$ mm). The second complementary band then almost perfectly compensates for these phase distortions to give rise to a nearly cylindrical transmitted wave-front ($x>20$ mm). The latter one is associated with a virtual source shifted by a distance $2L$ compared the \textit{real} source location. It thus appears to be inside the complementary bands to a downstream observer. At the output of the device, wave propagates as if the space containing the complementary bands had disappeared. This effect is highlighted by juxtaposing the wave-fronts in the areas upstream and downstream to the complementary bands (Fig.~\ref{fig:comp}c).

However, as shown by Fig.~\ref{fig:comp}b, the magnitude of the wave-field is not perfectly homogeneous at the output. A shaded area occurs in the superior part of the plate for $y \in [-50;-25]$ mm. A first reason is the decrease of the conversion coefficient between the forward $S_2$ and backward $S_{2b}$ modes for large angles of incidence~\cite{legrand2018negative}. In addition, a fraction of the incident wave is reflected by the corners between the positive and negative index areas (see for instance the red arrow in Fig.~\ref{fig:comp}b). These spurious reflections induce an heterogeneous angular distribution of the transmitted wave-front across the complementary bands. This effect had already been noticed by Pendry~\cite{pendrycomp} but, in an ideal case, those spurious reflections would be eliminated by destructive interferences between complementary corners via an evanescent coupling between them. In the present case, such a mechanism does not occur for Lamb waves and the reflections induced by the complementary corners give rise to a shadow zone behind the complementary bands.

Following this numerical study, the complementary band plate is manufactured on a duralumin plate by electrical discharge machining. 
The source is a piezoelectric transducer glued on the plate at $25$ mm from the complementary bands. The normal displacement is measured at the plate surface using a {photorefractive} interferometer  (see Methods of the accompanying paper). The wave-field filtered at the crossing frequency and its phase are shown in Fig.~\ref{fig:comp}g and h, respectively. The crossing frequency $f_c$ is here of $3.37$ MHz and slightly differs from its theoretical value ($f_c=3.32$ MHz). Negative index areas actually display a  thickness of 0.89 mm instead of 0.9 mm. The recorded wave-field shows some difference compared to its numerical counterpart (Figs.~\ref{fig:comp}d and e). Spurious reflections induced by thickness step corners, already highlighted by the numerical simulations in Fig.~\ref{fig:comp}d, are here enhanced by an imperfect tailoring of such singularities. The angular distribution of the transmitted wave-field magnitude thus exhibits strong flutuations. Nevertheless, the effect of the complementary bands is nicely highlighted by investigating the phase of the wave-field (Fig.~\ref{fig:comp}e). The juxtaposition of the wave-fronts at input and output of the complementary device displayed in Fig.~\ref{fig:comp}f shows the mutual compensation of the phase accumulated by the wave in each complementary band. 

The experimental implementation of these complementary bands are a first step towards the ability of cloaking a region of space by its anti-object demonstrated in the accompanying paper (Fig.~3). This experiment highlights the importance of a precise tailoring of negative corners. Despite their strong dependence on manufacturing imperfections, such singularities can be leveraged for wave trapping. This is illustrated by the experimental implementation of the double negative corner in the accompanying paper (Figs.~4 and 5).

\section*{{Design of the metalens}}

\begin{figure*}[t]
  \centering
  \includegraphics[width=\textwidth]{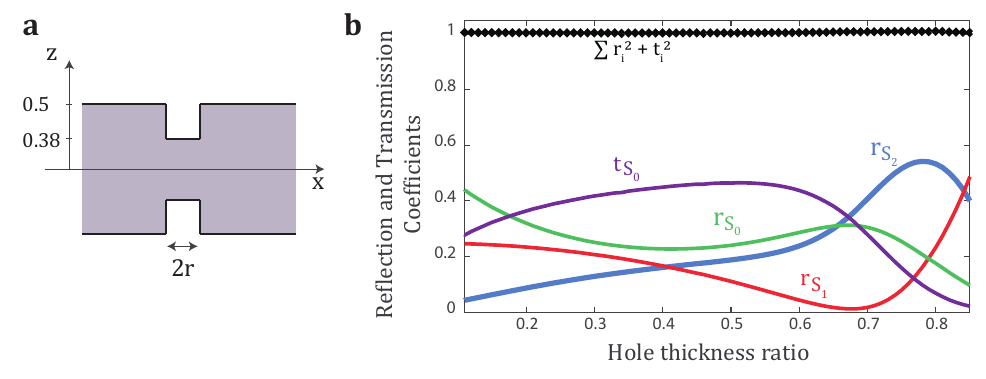}
  \caption{\textbf{Design of the metalens}. \textbf{a} Geometry of a single blind hole. \textbf{b} Theoretical reflection and transmission coefficients of the incident $S_2$ mode as a function of the thickness ratio in a 1 mm thick duralumin plate at the frequency $f=3.31$ MHz. }\label{fig:suppl:SupalensHoles}
\end{figure*}

The metalens used in the superlens simulation is made of blind holes~\cite{xeridat2011etude,dubois2014controle,Dubois2017} \al{that are known as efficient sub-wavelength resonators for Lamb waves}. Their geometry displayed in Fig.~\ref{fig:suppl:SupalensHoles}a has been designed to ensure various \al{properties}. First, the holes are symmetrical to avoid any conversion into anti-symmetric Lamb modes. Secondly, their radius and the distance between them are set at $2$ mm and 1 mm, respectively. These values are small compared to the $S_2$-mode wavelength ($\lambda_{c}=12.5$ mm) in order to efficiently scatter the evanescent components of the source. Finally, the thickness ratio of the blind holes has been fixed to maximize the reflection of the $S_{2}$-mode into itself. To that aim, we use the semi-analytical model \al{previously developed to study negative refraction at a thickness step~\cite{legrand2018negative}}.  Figure~\ref{fig:suppl:SupalensHoles}b shows the reflection and transmission coefficient of the incident $S_{2}$ mode as a function of the thickness ratio displayed by the blind hole. A maximum reflection coefficient is obtained for a ratio of $0.76$. Note that this semi-analytical model is valid for a plane interface. Its application to holes whose dimensions are lower than the wavelength is highly questionable. Nevertheless, the obtained value leads to a significant amount of scattering of the incident $S_2$ mode into itself by the blind holes (see Fig.~6 of the accompanying paper). 

\clearpage

{\noindent {\large \textbf{References}}}

\bibliographystyle{apsrev4-2}

\begin{thebibliography}{74}%
\makeatletter
\providecommand \@ifxundefined [1]{%
 \@ifx{#1\undefined}
}%
\providecommand \@ifnum [1]{%
 \ifnum #1\expandafter \@firstoftwo
 \else \expandafter \@secondoftwo
 \fi
}%
\providecommand \@ifx [1]{%
 \ifx #1\expandafter \@firstoftwo
 \else \expandafter \@secondoftwo
 \fi
}%
\providecommand \natexlab [1]{#1}%
\providecommand \enquote  [1]{``#1''}%
\providecommand \bibnamefont  [1]{#1}%
\providecommand \bibfnamefont [1]{#1}%
\providecommand \citenamefont [1]{#1}%
\providecommand \href@noop [0]{\@secondoftwo}%
\providecommand \href [0]{\begingroup \@sanitize@url \@href}%
\providecommand \@href[1]{\@@startlink{#1}\@@href}%
\providecommand \@@href[1]{\endgroup#1\@@endlink}%
\providecommand \@sanitize@url [0]{\catcode `\\12\catcode `\$12\catcode
  `\&12\catcode `\#12\catcode `\^12\catcode `\_12\catcode `\%12\relax}%
\providecommand \@@startlink[1]{}%
\providecommand \@@endlink[0]{}%
\providecommand \url  [0]{\begingroup\@sanitize@url \@url }%
\providecommand \@url [1]{\endgroup\@href {#1}{\urlprefix }}%
\providecommand \urlprefix  [0]{URL }%
\providecommand \Eprint [0]{\href }%
\providecommand \doibase [0]{https://doi.org/}%
\providecommand \selectlanguage [0]{\@gobble}%
\providecommand \bibinfo  [0]{\@secondoftwo}%
\providecommand \bibfield  [0]{\@secondoftwo}%
\providecommand \translation [1]{[#1]}%
\providecommand \BibitemOpen [0]{}%
\providecommand \bibitemStop [0]{}%
\providecommand \bibitemNoStop [0]{.\EOS\space}%
\providecommand \EOS [0]{\spacefactor3000\relax}%
\providecommand \BibitemShut  [1]{\csname bibitem#1\endcsname}%
\let\auto@bib@innerbib\@empty
\bibitem [{\citenamefont {Veselago}(1968)}]{veselago1968}%
  \BibitemOpen
  \bibfield  {author} {\bibinfo {author} {\bibfnamefont {V.~G.}\ \bibnamefont
  {Veselago}},\ }\href@noop {} {\bibfield  {journal} {\bibinfo  {journal}
  {Phys. Usp.}\ }\textbf {\bibinfo {volume} {10}},\ \bibinfo {pages} {509}
  (\bibinfo {year} {1968})}\BibitemShut {NoStop}%
\bibitem [{\citenamefont {Pendry}(2000)}]{pendry2000negative}%
  \BibitemOpen
  \bibfield  {author} {\bibinfo {author} {\bibfnamefont {J.~B.}\ \bibnamefont
  {Pendry}},\ }\href@noop {} {\bibfield  {journal} {\bibinfo  {journal} {Phys.
  Rev. Lett.}\ }\textbf {\bibinfo {volume} {85}},\ \bibinfo {pages} {3966}
  (\bibinfo {year} {2000})}\BibitemShut {NoStop}%
\bibitem [{\citenamefont {Pendry}(2004)}]{pendrycomp}%
  \BibitemOpen
  \bibfield  {author} {\bibinfo {author} {\bibfnamefont {J.~B.}\ \bibnamefont
  {Pendry}},\ }\href@noop {} {\bibfield  {journal} {\bibinfo  {journal}
  {Contemp. Phys.}\ }\textbf {\bibinfo {volume} {45}},\ \bibinfo {pages} {191}
  (\bibinfo {year} {2004})}\BibitemShut {NoStop}%
\bibitem [{\citenamefont {Leonhardt}(2006)}]{leonhardt2006optical}%
  \BibitemOpen
  \bibfield  {author} {\bibinfo {author} {\bibfnamefont {U.}~\bibnamefont
  {Leonhardt}},\ }\href@noop {} {\bibfield  {journal} {\bibinfo  {journal}
  {Science}\ }\textbf {\bibinfo {volume} {312}},\ \bibinfo {pages} {1777}
  (\bibinfo {year} {2006})}\BibitemShut {NoStop}%
\bibitem [{\citenamefont {Fang}\ \emph {et~al.}(2005)\citenamefont {Fang},
  \citenamefont {Lee}, \citenamefont {Sun},\ and\ \citenamefont
  {Zhang}}]{Fang2005}%
  \BibitemOpen
  \bibfield  {author} {\bibinfo {author} {\bibfnamefont {N.}~\bibnamefont
  {Fang}}, \bibinfo {author} {\bibfnamefont {H.}~\bibnamefont {Lee}}, \bibinfo
  {author} {\bibfnamefont {C.}~\bibnamefont {Sun}},\ and\ \bibinfo {author}
  {\bibfnamefont {X.}~\bibnamefont {Zhang}},\ }\href@noop {} {\bibfield
  {journal} {\bibinfo  {journal} {Science}\ }\textbf {\bibinfo {volume}
  {308}},\ \bibinfo {pages} {534} (\bibinfo {year} {2005})}\BibitemShut
  {NoStop}%
\bibitem [{\citenamefont {Liu}\ \emph {et~al.}(2003)\citenamefont {Liu},
  \citenamefont {Fang}, \citenamefont {Yen},\ and\ \citenamefont
  {Zhang}}]{liu2003rapid}%
  \BibitemOpen
  \bibfield  {author} {\bibinfo {author} {\bibfnamefont {Z.}~\bibnamefont
  {Liu}}, \bibinfo {author} {\bibfnamefont {N.}~\bibnamefont {Fang}}, \bibinfo
  {author} {\bibfnamefont {T.-J.}\ \bibnamefont {Yen}},\ and\ \bibinfo {author}
  {\bibfnamefont {X.}~\bibnamefont {Zhang}},\ }\href@noop {} {\bibfield
  {journal} {\bibinfo  {journal} {Appl. Phys. Lett.}\ }\textbf {\bibinfo
  {volume} {83}},\ \bibinfo {pages} {5184} (\bibinfo {year}
  {2003})}\BibitemShut {NoStop}%
\bibitem [{\citenamefont {Hu}\ \emph {et~al.}(2004)\citenamefont {Hu},
  \citenamefont {Shen}, \citenamefont {Liu}, \citenamefont {Fu},\ and\
  \citenamefont {Zi}}]{Hu2004}%
  \BibitemOpen
  \bibfield  {author} {\bibinfo {author} {\bibfnamefont {X.}~\bibnamefont
  {Hu}}, \bibinfo {author} {\bibfnamefont {Y.}~\bibnamefont {Shen}}, \bibinfo
  {author} {\bibfnamefont {X.}~\bibnamefont {Liu}}, \bibinfo {author}
  {\bibfnamefont {R.}~\bibnamefont {Fu}},\ and\ \bibinfo {author}
  {\bibfnamefont {J.}~\bibnamefont {Zi}},\ }\href@noop {} {\bibfield  {journal}
  {\bibinfo  {journal} {Phys. Rev. E}\ }\textbf {\bibinfo {volume} {69}},\
  \bibinfo {pages} {030201} (\bibinfo {year} {2004})}\BibitemShut {NoStop}%
\bibitem [{\citenamefont {Taubner}\ \emph {et~al.}(2006)\citenamefont
  {Taubner}, \citenamefont {Korobkin}, \citenamefont {Urzhumov}, \citenamefont
  {Shvets},\ and\ \citenamefont {Hillenbrand}}]{taubner2006near}%
  \BibitemOpen
  \bibfield  {author} {\bibinfo {author} {\bibfnamefont {T.}~\bibnamefont
  {Taubner}}, \bibinfo {author} {\bibfnamefont {D.}~\bibnamefont {Korobkin}},
  \bibinfo {author} {\bibfnamefont {Y.}~\bibnamefont {Urzhumov}}, \bibinfo
  {author} {\bibfnamefont {G.}~\bibnamefont {Shvets}},\ and\ \bibinfo {author}
  {\bibfnamefont {R.}~\bibnamefont {Hillenbrand}},\ }\href@noop {} {\bibfield
  {journal} {\bibinfo  {journal} {Science}\ }\textbf {\bibinfo {volume}
  {313}},\ \bibinfo {pages} {1595} (\bibinfo {year} {2006})}\BibitemShut
  {NoStop}%
\bibitem [{\citenamefont {Zhang}\ and\ \citenamefont
  {Liu}(2008)}]{zhang2008superlenses}%
  \BibitemOpen
  \bibfield  {author} {\bibinfo {author} {\bibfnamefont {X.}~\bibnamefont
  {Zhang}}\ and\ \bibinfo {author} {\bibfnamefont {Z.}~\bibnamefont {Liu}},\
  }\href@noop {} {\bibfield  {journal} {\bibinfo  {journal} {Nat. Mater.}\
  }\textbf {\bibinfo {volume} {7}},\ \bibinfo {pages} {435} (\bibinfo {year}
  {2008})}\BibitemShut {NoStop}%
\bibitem [{\citenamefont {Sukhovich}\ \emph {et~al.}(2009)\citenamefont
  {Sukhovich}, \citenamefont {Merheb}, \citenamefont {Muralidharan},
  \citenamefont {Vasseur}, \citenamefont {Pennec}, \citenamefont {Deymier},\
  and\ \citenamefont {Page}}]{Sukhovich2009}%
  \BibitemOpen
  \bibfield  {author} {\bibinfo {author} {\bibfnamefont {A.}~\bibnamefont
  {Sukhovich}}, \bibinfo {author} {\bibfnamefont {B.}~\bibnamefont {Merheb}},
  \bibinfo {author} {\bibfnamefont {K.}~\bibnamefont {Muralidharan}}, \bibinfo
  {author} {\bibfnamefont {J.~O.}\ \bibnamefont {Vasseur}}, \bibinfo {author}
  {\bibfnamefont {Y.}~\bibnamefont {Pennec}}, \bibinfo {author} {\bibfnamefont
  {P.~A.}\ \bibnamefont {Deymier}},\ and\ \bibinfo {author} {\bibfnamefont
  {J.~H.}\ \bibnamefont {Page}},\ }\href@noop {} {\bibfield  {journal}
  {\bibinfo  {journal} {Phys. Rev. Lett.}\ }\textbf {\bibinfo {volume} {102}},\
  \bibinfo {pages} {154301} (\bibinfo {year} {2009})}\BibitemShut {NoStop}%
\bibitem [{\citenamefont {Zhu}\ \emph {et~al.}(2014)\citenamefont {Zhu},
  \citenamefont {Liu}, \citenamefont {Hu}, \citenamefont {Sun},\ and\
  \citenamefont {Huang}}]{zhu2014negative}%
  \BibitemOpen
  \bibfield  {author} {\bibinfo {author} {\bibfnamefont {R.}~\bibnamefont
  {Zhu}}, \bibinfo {author} {\bibfnamefont {X.}~\bibnamefont {Liu}}, \bibinfo
  {author} {\bibfnamefont {G.}~\bibnamefont {Hu}}, \bibinfo {author}
  {\bibfnamefont {C.}~\bibnamefont {Sun}},\ and\ \bibinfo {author}
  {\bibfnamefont {G.}~\bibnamefont {Huang}},\ }\href@noop {} {\bibfield
  {journal} {\bibinfo  {journal} {Nat. Commun.}\ }\textbf {\bibinfo {volume}
  {5}},\ \bibinfo {pages} {5510} (\bibinfo {year} {2014})}\BibitemShut
  {NoStop}%
\bibitem [{\citenamefont {Park}\ \emph {et~al.}(2015)\citenamefont {Park},
  \citenamefont {Park}, \citenamefont {Lee},\ and\ \citenamefont
  {Lee}}]{Park2015}%
  \BibitemOpen
  \bibfield  {author} {\bibinfo {author} {\bibfnamefont {J.~J.}\ \bibnamefont
  {Park}}, \bibinfo {author} {\bibfnamefont {C.~M.}\ \bibnamefont {Park}},
  \bibinfo {author} {\bibfnamefont {K.~J.~B.}\ \bibnamefont {Lee}},\ and\
  \bibinfo {author} {\bibfnamefont {S.~H.}\ \bibnamefont {Lee}},\ }\href@noop
  {} {\bibfield  {journal} {\bibinfo  {journal} {Appl. Phys. Lett.}\ }\textbf
  {\bibinfo {volume} {106}},\ \bibinfo {pages} {051901} (\bibinfo {year}
  {2015})}\BibitemShut {NoStop}%
\bibitem [{\citenamefont {Brunet}\ \emph {et~al.}(2015)\citenamefont {Brunet},
  \citenamefont {Merlin}, \citenamefont {Mascaro}, \citenamefont {Zimny},
  \citenamefont {Leng}, \citenamefont {Poncelet}, \citenamefont {Aristégui},\
  and\ \citenamefont {Mondain-Monval}}]{Brunet2015}%
  \BibitemOpen
  \bibfield  {author} {\bibinfo {author} {\bibfnamefont {T.}~\bibnamefont
  {Brunet}}, \bibinfo {author} {\bibfnamefont {A.}~\bibnamefont {Merlin}},
  \bibinfo {author} {\bibfnamefont {B.}~\bibnamefont {Mascaro}}, \bibinfo
  {author} {\bibfnamefont {K.}~\bibnamefont {Zimny}}, \bibinfo {author}
  {\bibfnamefont {J.}~\bibnamefont {Leng}}, \bibinfo {author} {\bibfnamefont
  {O.}~\bibnamefont {Poncelet}}, \bibinfo {author} {\bibfnamefont
  {C.}~\bibnamefont {Aristégui}},\ and\ \bibinfo {author} {\bibfnamefont
  {O.}~\bibnamefont {Mondain-Monval}},\ }\href@noop {} {\bibfield  {journal}
  {\bibinfo  {journal} {Nat. Mater.}\ }\textbf {\bibinfo {volume} {14}},\
  \bibinfo {pages} {374} (\bibinfo {year} {2015})}\BibitemShut {NoStop}%
\bibitem [{\citenamefont {Kaina}\ \emph {et~al.}(2015)\citenamefont {Kaina},
  \citenamefont {Lemoult}, \citenamefont {Fink},\ and\ \citenamefont
  {Lerosey}}]{Kaina2015}%
  \BibitemOpen
  \bibfield  {author} {\bibinfo {author} {\bibfnamefont {N.}~\bibnamefont
  {Kaina}}, \bibinfo {author} {\bibfnamefont {F.}~\bibnamefont {Lemoult}},
  \bibinfo {author} {\bibfnamefont {M.}~\bibnamefont {Fink}},\ and\ \bibinfo
  {author} {\bibfnamefont {G.}~\bibnamefont {Lerosey}},\ }\href@noop {}
  {\bibfield  {journal} {\bibinfo  {journal} {Nature}\ }\textbf {\bibinfo
  {volume} {525}},\ \bibinfo {pages} {77} (\bibinfo {year} {2015})}\BibitemShut
  {NoStop}%
\bibitem [{\citenamefont {Xu}\ \emph {et~al.}(2017)\citenamefont {Xu},
  \citenamefont {Fang}, \citenamefont {Jia}, \citenamefont {Ji},\ and\
  \citenamefont {Hang}}]{xu2017realization}%
  \BibitemOpen
  \bibfield  {author} {\bibinfo {author} {\bibfnamefont {T.}~\bibnamefont
  {Xu}}, \bibinfo {author} {\bibfnamefont {A.}~\bibnamefont {Fang}}, \bibinfo
  {author} {\bibfnamefont {Z.}~\bibnamefont {Jia}}, \bibinfo {author}
  {\bibfnamefont {L.}~\bibnamefont {Ji}},\ and\ \bibinfo {author}
  {\bibfnamefont {Z.~H.}\ \bibnamefont {Hang}},\ }\href@noop {} {\bibfield
  {journal} {\bibinfo  {journal} {Opt. Lett.}\ }\textbf {\bibinfo {volume}
  {42}},\ \bibinfo {pages} {4909} (\bibinfo {year} {2017})}\BibitemShut
  {NoStop}%
\bibitem [{\citenamefont {Smith}\ \emph {et~al.}(2003)\citenamefont {Smith},
  \citenamefont {Schurig}, \citenamefont {Rosenbluth}, \citenamefont {Schultz},
  \citenamefont {Ramakrishna},\ and\ \citenamefont {Pendry}}]{Smith2003}%
  \BibitemOpen
  \bibfield  {author} {\bibinfo {author} {\bibfnamefont {D.~R.}\ \bibnamefont
  {Smith}}, \bibinfo {author} {\bibfnamefont {D.}~\bibnamefont {Schurig}},
  \bibinfo {author} {\bibfnamefont {M.}~\bibnamefont {Rosenbluth}}, \bibinfo
  {author} {\bibfnamefont {S.}~\bibnamefont {Schultz}}, \bibinfo {author}
  {\bibfnamefont {S.~A.}\ \bibnamefont {Ramakrishna}},\ and\ \bibinfo {author}
  {\bibfnamefont {J.~B.}\ \bibnamefont {Pendry}},\ }\href@noop {} {\bibfield
  {journal} {\bibinfo  {journal} {Appl. Phys. Lett.}\ }\textbf {\bibinfo
  {volume} {82}},\ \bibinfo {pages} {1506} (\bibinfo {year}
  {2003})}\BibitemShut {NoStop}%
\bibitem [{\citenamefont {Tolstoy}\ and\ \citenamefont
  {Usdin}(1957)}]{tolstoy1957wave}%
  \BibitemOpen
  \bibfield  {author} {\bibinfo {author} {\bibfnamefont {I.}~\bibnamefont
  {Tolstoy}}\ and\ \bibinfo {author} {\bibfnamefont {E.}~\bibnamefont
  {Usdin}},\ }\href@noop {} {\bibfield  {journal} {\bibinfo  {journal} {J.
  Acoust. Soc. Am.}\ }\textbf {\bibinfo {volume} {29}},\ \bibinfo {pages} {37}
  (\bibinfo {year} {1957})}\BibitemShut {NoStop}%
\bibitem [{\citenamefont {Prada}\ \emph {et~al.}(2005)\citenamefont {Prada},
  \citenamefont {Balogun},\ and\ \citenamefont {Murray}}]{prada2005laser}%
  \BibitemOpen
  \bibfield  {author} {\bibinfo {author} {\bibfnamefont {C.}~\bibnamefont
  {Prada}}, \bibinfo {author} {\bibfnamefont {O.}~\bibnamefont {Balogun}},\
  and\ \bibinfo {author} {\bibfnamefont {T.~W.}\ \bibnamefont {Murray}},\
  }\href@noop {} {\bibfield  {journal} {\bibinfo  {journal} {Appl. Phys.
  Lett.}\ }\textbf {\bibinfo {volume} {87}},\ \bibinfo {pages} {194109}
  (\bibinfo {year} {2005})}\BibitemShut {NoStop}%
\bibitem [{\citenamefont {Holland}\ and\ \citenamefont
  {Chimenti}(2003)}]{holland2003air}%
  \BibitemOpen
  \bibfield  {author} {\bibinfo {author} {\bibfnamefont {S.~D.}\ \bibnamefont
  {Holland}}\ and\ \bibinfo {author} {\bibfnamefont {D.~E.}\ \bibnamefont
  {Chimenti}},\ }\href@noop {} {\bibfield  {journal} {\bibinfo  {journal}
  {Appl. Phys. Lett.}\ }\textbf {\bibinfo {volume} {83}},\ \bibinfo {pages}
  {2704} (\bibinfo {year} {2003})}\BibitemShut {NoStop}%
\bibitem [{\citenamefont {Xie}\ \emph {et~al.}(2019)\citenamefont {Xie},
  \citenamefont {Mezil}, \citenamefont {Otsuka}, \citenamefont {Tomoda},
  \citenamefont {Laurent}, \citenamefont {Matsuda}, \citenamefont {Shen},\ and\
  \citenamefont {Wright}}]{Xie2019}%
  \BibitemOpen
  \bibfield  {author} {\bibinfo {author} {\bibfnamefont {Q.}~\bibnamefont
  {Xie}}, \bibinfo {author} {\bibfnamefont {S.}~\bibnamefont {Mezil}}, \bibinfo
  {author} {\bibfnamefont {P.~H.}\ \bibnamefont {Otsuka}}, \bibinfo {author}
  {\bibfnamefont {M.}~\bibnamefont {Tomoda}}, \bibinfo {author} {\bibfnamefont
  {J.}~\bibnamefont {Laurent}}, \bibinfo {author} {\bibfnamefont
  {O.}~\bibnamefont {Matsuda}}, \bibinfo {author} {\bibfnamefont
  {Z.}~\bibnamefont {Shen}},\ and\ \bibinfo {author} {\bibfnamefont {O.~B.}\
  \bibnamefont {Wright}},\ }\href@noop {} {\bibfield  {journal} {\bibinfo
  {journal} {Nat. Commun.}\ }\textbf {\bibinfo {volume} {10}},\ \bibinfo
  {pages} {2228} (\bibinfo {year} {2019})}\BibitemShut {NoStop}%
\bibitem [{\citenamefont {Bramhavar}\ \emph {et~al.}(2011)\citenamefont
  {Bramhavar}, \citenamefont {Prada}, \citenamefont {Maznev}, \citenamefont
  {Every}, \citenamefont {Norris},\ and\ \citenamefont
  {Murray}}]{bramhavar2011negative}%
  \BibitemOpen
  \bibfield  {author} {\bibinfo {author} {\bibfnamefont {S.}~\bibnamefont
  {Bramhavar}}, \bibinfo {author} {\bibfnamefont {C.}~\bibnamefont {Prada}},
  \bibinfo {author} {\bibfnamefont {A.~A.}\ \bibnamefont {Maznev}}, \bibinfo
  {author} {\bibfnamefont {A.~G.}\ \bibnamefont {Every}}, \bibinfo {author}
  {\bibfnamefont {T.~B.}\ \bibnamefont {Norris}},\ and\ \bibinfo {author}
  {\bibfnamefont {T.~W.}\ \bibnamefont {Murray}},\ }\href@noop {} {\bibfield
  {journal} {\bibinfo  {journal} {Phys. Rev. B}\ }\textbf {\bibinfo {volume}
  {83}},\ \bibinfo {pages} {014106} (\bibinfo {year} {2011})}\BibitemShut
  {NoStop}%
\bibitem [{\citenamefont {Philippe}\ \emph {et~al.}(2015)\citenamefont
  {Philippe}, \citenamefont {Murray},\ and\ \citenamefont
  {Prada}}]{philippe2015focusing}%
  \BibitemOpen
  \bibfield  {author} {\bibinfo {author} {\bibfnamefont {F.~D.}\ \bibnamefont
  {Philippe}}, \bibinfo {author} {\bibfnamefont {T.~W.}\ \bibnamefont
  {Murray}},\ and\ \bibinfo {author} {\bibfnamefont {C.}~\bibnamefont
  {Prada}},\ }\href@noop {} {\bibfield  {journal} {\bibinfo  {journal} {Sci.
  Rep.}\ }\textbf {\bibinfo {volume} {5}},\ \bibinfo {pages} {11112} (\bibinfo
  {year} {2015})}\BibitemShut {NoStop}%
\bibitem [{\citenamefont {Manjunath}\ and\ \citenamefont
  {Rajagopal}(2019)}]{Manjunath2019}%
  \BibitemOpen
  \bibfield  {author} {\bibinfo {author} {\bibfnamefont {C.~T.}\ \bibnamefont
  {Manjunath}}\ and\ \bibinfo {author} {\bibfnamefont {P.}~\bibnamefont
  {Rajagopal}},\ }\href@noop {} {\bibfield  {journal} {\bibinfo  {journal}
  {Sci. Rep.}\ }\textbf {\bibinfo {volume} {9}},\ \bibinfo {pages} {6368}
  (\bibinfo {year} {2019})}\BibitemShut {NoStop}%
\bibitem [{\citenamefont {G{\'e}rardin}\ \emph {et~al.}(2016)\citenamefont
  {G{\'e}rardin}, \citenamefont {Laurent}, \citenamefont {Prada},\ and\
  \citenamefont {Aubry}}]{gerardin2016negative}%
  \BibitemOpen
  \bibfield  {author} {\bibinfo {author} {\bibfnamefont {B.}~\bibnamefont
  {G{\'e}rardin}}, \bibinfo {author} {\bibfnamefont {J.}~\bibnamefont
  {Laurent}}, \bibinfo {author} {\bibfnamefont {C.}~\bibnamefont {Prada}},\
  and\ \bibinfo {author} {\bibfnamefont {A.}~\bibnamefont {Aubry}},\
  }\href@noop {} {\bibfield  {journal} {\bibinfo  {journal} {J. Acoust. Soc.
  Am.}\ }\textbf {\bibinfo {volume} {140}},\ \bibinfo {pages} {591} (\bibinfo
  {year} {2016})}\BibitemShut {NoStop}%
\bibitem [{\citenamefont {Veres}\ \emph {et~al.}(2016)\citenamefont {Veres},
  \citenamefont {Gr{\"e}unsteidl}, \citenamefont {Stobbe},\ and\ \citenamefont
  {Murray}}]{veres2016broad}%
  \BibitemOpen
  \bibfield  {author} {\bibinfo {author} {\bibfnamefont {I.~A.}\ \bibnamefont
  {Veres}}, \bibinfo {author} {\bibfnamefont {C.}~\bibnamefont
  {Gr{\"e}unsteidl}}, \bibinfo {author} {\bibfnamefont {D.~M.}\ \bibnamefont
  {Stobbe}},\ and\ \bibinfo {author} {\bibfnamefont {T.~W.}\ \bibnamefont
  {Murray}},\ }\href@noop {} {\bibfield  {journal} {\bibinfo  {journal} {Phys.
  Rev. B}\ }\textbf {\bibinfo {volume} {93}},\ \bibinfo {pages} {174304}
  (\bibinfo {year} {2016})}\BibitemShut {NoStop}%
\bibitem [{\citenamefont {G{\'e}rardin}\ \emph {et~al.}(2019)\citenamefont
  {G{\'e}rardin}, \citenamefont {Laurent}, \citenamefont {Legrand},
  \citenamefont {Prada},\ and\ \citenamefont {Aubry}}]{gerardin2019negative}%
  \BibitemOpen
  \bibfield  {author} {\bibinfo {author} {\bibfnamefont {B.}~\bibnamefont
  {G{\'e}rardin}}, \bibinfo {author} {\bibfnamefont {J.}~\bibnamefont
  {Laurent}}, \bibinfo {author} {\bibfnamefont {F.}~\bibnamefont {Legrand}},
  \bibinfo {author} {\bibfnamefont {C.}~\bibnamefont {Prada}},\ and\ \bibinfo
  {author} {\bibfnamefont {A.}~\bibnamefont {Aubry}},\ }\href@noop {}
  {\bibfield  {journal} {\bibinfo  {journal} {Sci. Rep.}\ }\textbf {\bibinfo
  {volume} {9}},\ \bibinfo {pages} {2135} (\bibinfo {year} {2019})}\BibitemShut
  {NoStop}%
\bibitem [{\citenamefont {Legrand}\ \emph {et~al.}(2018)\citenamefont
  {Legrand}, \citenamefont {G{\'e}rardin}, \citenamefont {Laurent},
  \citenamefont {Prada},\ and\ \citenamefont {Aubry}}]{legrand2018negative}%
  \BibitemOpen
  \bibfield  {author} {\bibinfo {author} {\bibfnamefont {F.}~\bibnamefont
  {Legrand}}, \bibinfo {author} {\bibfnamefont {B.}~\bibnamefont
  {G{\'e}rardin}}, \bibinfo {author} {\bibfnamefont {J.}~\bibnamefont
  {Laurent}}, \bibinfo {author} {\bibfnamefont {C.}~\bibnamefont {Prada}},\
  and\ \bibinfo {author} {\bibfnamefont {A.}~\bibnamefont {Aubry}},\
  }\href@noop {} {\bibfield  {journal} {\bibinfo  {journal} {Phys. Rev. B}\
  }\textbf {\bibinfo {volume} {98}},\ \bibinfo {pages} {214114} (\bibinfo
  {year} {2018})}\BibitemShut {NoStop}%
\bibitem [{\citenamefont {Lai}\ \emph {et~al.}(2009)\citenamefont {Lai},
  \citenamefont {Chen}, \citenamefont {Zhang},\ and\ \citenamefont
  {Chan}}]{lai2009complementary}%
  \BibitemOpen
  \bibfield  {author} {\bibinfo {author} {\bibfnamefont {Y.}~\bibnamefont
  {Lai}}, \bibinfo {author} {\bibfnamefont {H.}~\bibnamefont {Chen}}, \bibinfo
  {author} {\bibfnamefont {Z.-Q.}\ \bibnamefont {Zhang}},\ and\ \bibinfo
  {author} {\bibfnamefont {C.}~\bibnamefont {Chan}},\ }\href@noop {} {\bibfield
   {journal} {\bibinfo  {journal} {Phys. Rev. Lett.}\ }\textbf {\bibinfo
  {volume} {102}},\ \bibinfo {pages} {093901} (\bibinfo {year}
  {2009})}\BibitemShut {NoStop}%
\bibitem [{\citenamefont {Nguy\^{e}n}(2016)}]{Nguyen}%
  \BibitemOpen
  \bibfield  {author} {\bibinfo {author} {\bibfnamefont {L.~H.}\ \bibnamefont
  {Nguy\^{e}n}},\ }\href@noop {} {\bibfield  {journal} {\bibinfo  {journal}
  {Ann. Inst. H. Poincaré Anal. NonLin\'{e}aire}\ }\textbf {\bibinfo {volume}
  {33}},\ \bibinfo {pages} {1509} (\bibinfo {year} {2016})}\BibitemShut
  {NoStop}%
\bibitem [{\citenamefont {Nguyen}(2019)}]{Nguyen2019}%
  \BibitemOpen
  \bibfield  {author} {\bibinfo {author} {\bibfnamefont {H.-M.}\ \bibnamefont
  {Nguyen}},\ }\href@noop {} {\bibfield  {journal} {\bibinfo  {journal} {ESAIM:
  COCV}\ }\textbf {\bibinfo {volume} {25}},\ \bibinfo {pages} {29} (\bibinfo
  {year} {2019})}\BibitemShut {NoStop}%
\bibitem [{\citenamefont {Notomi}(2002)}]{notomi2002negative}%
  \BibitemOpen
  \bibfield  {author} {\bibinfo {author} {\bibfnamefont {M.}~\bibnamefont
  {Notomi}},\ }\href@noop {} {\bibfield  {journal} {\bibinfo  {journal} {Opt
  Quantum Electron.}\ }\textbf {\bibinfo {volume} {34}},\ \bibinfo {pages}
  {133} (\bibinfo {year} {2002})}\BibitemShut {NoStop}%
\bibitem [{\citenamefont {Pendry}\ and\ \citenamefont
  {Ramakrishna}(2003)}]{pendry2003focusing}%
  \BibitemOpen
  \bibfield  {author} {\bibinfo {author} {\bibfnamefont {J.~B.}\ \bibnamefont
  {Pendry}}\ and\ \bibinfo {author} {\bibfnamefont {S.~A.}\ \bibnamefont
  {Ramakrishna}},\ }\href@noop {} {\bibfield  {journal} {\bibinfo  {journal}
  {J. Phys. Condens. Matter}\ }\textbf {\bibinfo {volume} {15}},\ \bibinfo
  {pages} {6345} (\bibinfo {year} {2003})}\BibitemShut {NoStop}%
\bibitem [{\citenamefont {Bossy}\ \emph {et~al.}(2004)\citenamefont {Bossy},
  \citenamefont {Talmant},\ and\ \citenamefont {Laugier}}]{bossy2004three}%
  \BibitemOpen
  \bibfield  {author} {\bibinfo {author} {\bibfnamefont {E.}~\bibnamefont
  {Bossy}}, \bibinfo {author} {\bibfnamefont {M.}~\bibnamefont {Talmant}},\
  and\ \bibinfo {author} {\bibfnamefont {P.}~\bibnamefont {Laugier}},\
  }\href@noop {} {\bibfield  {journal} {\bibinfo  {journal} {J. Acoust. Soc.
  Am.}\ }\textbf {\bibinfo {volume} {115}},\ \bibinfo {pages} {2314} (\bibinfo
  {year} {2004})}\BibitemShut {NoStop}%
\bibitem [{\citenamefont {Bossy}(2003)}]{simsonic}%
  \BibitemOpen
  \bibfield  {author} {\bibinfo {author} {\bibfnamefont {E.}~\bibnamefont
  {Bossy}},\ }\href@noop {} {\bibinfo {title} {Simsonic, a {FDTD} simulation
  freeware}} (\bibinfo {year} {2003}),\ \bibinfo {note}
  {www.simsonic.fr}\BibitemShut {NoStop}%
\bibitem [{\citenamefont {Mahajan}(1982)}]{mahajan1982strehl}%
  \BibitemOpen
  \bibfield  {author} {\bibinfo {author} {\bibfnamefont {V.~N.}\ \bibnamefont
  {Mahajan}},\ }\href@noop {} {\bibfield  {journal} {\bibinfo  {journal} {J.
  Opt. Soc. Am.}\ }\textbf {\bibinfo {volume} {72}},\ \bibinfo {pages} {1258}
  (\bibinfo {year} {1982})}\BibitemShut {NoStop}%
\bibitem [{\citenamefont {Milton}\ \emph {et~al.}(2005)\citenamefont {Milton},
  \citenamefont {Nicorovici}, \citenamefont {McPhedran},\ and\ \citenamefont
  {Podolskiy}}]{Milton2005}%
  \BibitemOpen
  \bibfield  {author} {\bibinfo {author} {\bibfnamefont {G.~W.}\ \bibnamefont
  {Milton}}, \bibinfo {author} {\bibfnamefont {N.-A.~P.}\ \bibnamefont
  {Nicorovici}}, \bibinfo {author} {\bibfnamefont {R.~C.}\ \bibnamefont
  {McPhedran}},\ and\ \bibinfo {author} {\bibfnamefont {V.~A.}\ \bibnamefont
  {Podolskiy}},\ }\href@noop {} {\bibfield  {journal} {\bibinfo  {journal}
  {Proc. R. Soc. A.}\ }\textbf {\bibinfo {volume} {461}},\ \bibinfo {pages}
  {3999} (\bibinfo {year} {2005})}\BibitemShut {NoStop}%
\bibitem [{\citenamefont {Milton}\ and\ \citenamefont
  {Nicorovici}(2006)}]{milton2006cloaking}%
  \BibitemOpen
  \bibfield  {author} {\bibinfo {author} {\bibfnamefont {G.~W.}\ \bibnamefont
  {Milton}}\ and\ \bibinfo {author} {\bibfnamefont {N.-A.~P.}\ \bibnamefont
  {Nicorovici}},\ }\href@noop {} {\bibfield  {journal} {\bibinfo  {journal}
  {Proc. Math. Phys. Eng. Sci.}\ }\textbf {\bibinfo {volume} {462}},\ \bibinfo
  {pages} {3027} (\bibinfo {year} {2006})}\BibitemShut {NoStop}%
\bibitem [{\citenamefont {McPhedran}\ and\ \citenamefont
  {Milton}(2020)}]{McPhedran2020}%
  \BibitemOpen
  \bibfield  {author} {\bibinfo {author} {\bibfnamefont {R.~C.}\ \bibnamefont
  {McPhedran}}\ and\ \bibinfo {author} {\bibfnamefont {G.~W.}\ \bibnamefont
  {Milton}},\ }\href@noop {} {\bibfield  {journal} {\bibinfo  {journal}
  {Comptes Rendus. Physique}\ }\textbf {\bibinfo {volume} {21}},\ \bibinfo
  {pages} {409} (\bibinfo {year} {2020})}\BibitemShut {NoStop}%
\bibitem [{\citenamefont {Deng}\ \emph {et~al.}(2020)\citenamefont {Deng},
  \citenamefont {Li},\ and\ \citenamefont {Liu}}]{Deng2020}%
  \BibitemOpen
  \bibfield  {author} {\bibinfo {author} {\bibfnamefont {Y.}~\bibnamefont
  {Deng}}, \bibinfo {author} {\bibfnamefont {H.}~\bibnamefont {Li}},\ and\
  \bibinfo {author} {\bibfnamefont {H.}~\bibnamefont {Liu}},\ }\href@noop {}
  {\bibfield  {journal} {\bibinfo  {journal} {J. Elast.}\ }\textbf {\bibinfo
  {volume} {140}},\ \bibinfo {pages} {213} (\bibinfo {year}
  {2020})}\BibitemShut {NoStop}%
\bibitem [{\citenamefont {Lemoult}\ \emph {et~al.}(2010)\citenamefont
  {Lemoult}, \citenamefont {Lerosey}, \citenamefont {de~Rosny},\ and\
  \citenamefont {Fink}}]{lemoult2010resonant}%
  \BibitemOpen
  \bibfield  {author} {\bibinfo {author} {\bibfnamefont {F.}~\bibnamefont
  {Lemoult}}, \bibinfo {author} {\bibfnamefont {G.}~\bibnamefont {Lerosey}},
  \bibinfo {author} {\bibfnamefont {J.}~\bibnamefont {de~Rosny}},\ and\
  \bibinfo {author} {\bibfnamefont {M.}~\bibnamefont {Fink}},\ }\href@noop {}
  {\bibfield  {journal} {\bibinfo  {journal} {Phys. Rev. Lett.}\ }\textbf
  {\bibinfo {volume} {104}},\ \bibinfo {pages} {203901} (\bibinfo {year}
  {2010})}\BibitemShut {NoStop}%
\bibitem [{\citenamefont {Lemoult}\ \emph {et~al.}(2011)\citenamefont
  {Lemoult}, \citenamefont {Fink},\ and\ \citenamefont
  {Lerosey}}]{lemoult2011acoustic}%
  \BibitemOpen
  \bibfield  {author} {\bibinfo {author} {\bibfnamefont {F.}~\bibnamefont
  {Lemoult}}, \bibinfo {author} {\bibfnamefont {M.}~\bibnamefont {Fink}},\ and\
  \bibinfo {author} {\bibfnamefont {G.}~\bibnamefont {Lerosey}},\ }\href@noop
  {} {\bibfield  {journal} {\bibinfo  {journal} {Phys. Rev. Lett.}\ }\textbf
  {\bibinfo {volume} {107}},\ \bibinfo {pages} {064301} (\bibinfo {year}
  {2011})}\BibitemShut {NoStop}%
\bibitem [{\citenamefont {Lemoult}\ \emph {et~al.}(2012)\citenamefont
  {Lemoult}, \citenamefont {Fink},\ and\ \citenamefont
  {Lerosey}}]{Lemoult2012}%
  \BibitemOpen
  \bibfield  {author} {\bibinfo {author} {\bibfnamefont {F.}~\bibnamefont
  {Lemoult}}, \bibinfo {author} {\bibfnamefont {M.}~\bibnamefont {Fink}},\ and\
  \bibinfo {author} {\bibfnamefont {G.}~\bibnamefont {Lerosey}},\ }\href@noop
  {} {\bibfield  {journal} {\bibinfo  {journal} {Nat. Commun.}\ }\textbf
  {\bibinfo {volume} {3}},\ \bibinfo {pages} {889} (\bibinfo {year}
  {2012})}\BibitemShut {NoStop}%
\bibitem [{\citenamefont {Meitzler}(1965)}]{meitzler1965backward}%
  \BibitemOpen
  \bibfield  {author} {\bibinfo {author} {\bibfnamefont {A.~H.}\ \bibnamefont
  {Meitzler}},\ }\href@noop {} {\bibfield  {journal} {\bibinfo  {journal} {J.
  Acoust. Soc. Am.}\ }\textbf {\bibinfo {volume} {38}},\ \bibinfo {pages} {835}
  (\bibinfo {year} {1965})}\BibitemShut {NoStop}%
\bibitem [{\citenamefont {Prada}\ \emph {et~al.}(2008)\citenamefont {Prada},
  \citenamefont {Clorennec},\ and\ \citenamefont {Royer}}]{Prada2008}%
  \BibitemOpen
  \bibfield  {author} {\bibinfo {author} {\bibfnamefont {C.}~\bibnamefont
  {Prada}}, \bibinfo {author} {\bibfnamefont {D.}~\bibnamefont {Clorennec}},\
  and\ \bibinfo {author} {\bibfnamefont {D.}~\bibnamefont {Royer}},\
  }\href@noop {} {\bibfield  {journal} {\bibinfo  {journal} {J. Acoust. Soc.
  Am.}\ }\textbf {\bibinfo {volume} {124}},\ \bibinfo {pages} {203} (\bibinfo
  {year} {2008})}\BibitemShut {NoStop}%
\bibitem [{\citenamefont {Stenger}\ \emph {et~al.}(2012)\citenamefont
  {Stenger}, \citenamefont {Wilhelm},\ and\ \citenamefont
  {Wegener}}]{stenger2012experiments}%
  \BibitemOpen
  \bibfield  {author} {\bibinfo {author} {\bibfnamefont {N.}~\bibnamefont
  {Stenger}}, \bibinfo {author} {\bibfnamefont {M.}~\bibnamefont {Wilhelm}},\
  and\ \bibinfo {author} {\bibfnamefont {M.}~\bibnamefont {Wegener}},\
  }\href@noop {} {\bibfield  {journal} {\bibinfo  {journal} {Phys. Rev. Lett.}\
  }\textbf {\bibinfo {volume} {108}},\ \bibinfo {pages} {014301} (\bibinfo
  {year} {2012})}\BibitemShut {NoStop}%
\bibitem [{\citenamefont {Jin}\ \emph {et~al.}(2016)\citenamefont {Jin},
  \citenamefont {Torrent}, \citenamefont {Pennec}, \citenamefont {Pan},\ and\
  \citenamefont {Djafari-Rouhani}}]{jin2016gradient}%
  \BibitemOpen
  \bibfield  {author} {\bibinfo {author} {\bibfnamefont {Y.}~\bibnamefont
  {Jin}}, \bibinfo {author} {\bibfnamefont {D.}~\bibnamefont {Torrent}},
  \bibinfo {author} {\bibfnamefont {Y.}~\bibnamefont {Pennec}}, \bibinfo
  {author} {\bibfnamefont {Y.}~\bibnamefont {Pan}},\ and\ \bibinfo {author}
  {\bibfnamefont {B.}~\bibnamefont {Djafari-Rouhani}},\ }\href@noop {}
  {\bibfield  {journal} {\bibinfo  {journal} {Sci. Rep.}\ }\textbf {\bibinfo
  {volume} {6}},\ \bibinfo {pages} {24437} (\bibinfo {year}
  {2016})}\BibitemShut {NoStop}%
\bibitem [{\citenamefont {Al{\`u}}\ and\ \citenamefont
  {Engheta}(2005)}]{alu2005achieving}%
  \BibitemOpen
  \bibfield  {author} {\bibinfo {author} {\bibfnamefont {A.}~\bibnamefont
  {Al{\`u}}}\ and\ \bibinfo {author} {\bibfnamefont {N.}~\bibnamefont
  {Engheta}},\ }\href@noop {} {\bibfield  {journal} {\bibinfo  {journal} {Phys.
  Rev. E.}\ }\textbf {\bibinfo {volume} {72}},\ \bibinfo {pages} {016623}
  (\bibinfo {year} {2005})}\BibitemShut {NoStop}%
\bibitem [{\citenamefont {Schurig}\ \emph {et~al.}(2006)\citenamefont
  {Schurig}, \citenamefont {Mock}, \citenamefont {Justice}, \citenamefont
  {Cummer}, \citenamefont {Pendry}, \citenamefont {Starr},\ and\ \citenamefont
  {Smith}}]{schurig2006metamaterial}%
  \BibitemOpen
  \bibfield  {author} {\bibinfo {author} {\bibfnamefont {D.}~\bibnamefont
  {Schurig}}, \bibinfo {author} {\bibfnamefont {J.~J.}\ \bibnamefont {Mock}},
  \bibinfo {author} {\bibfnamefont {B.}~\bibnamefont {Justice}}, \bibinfo
  {author} {\bibfnamefont {S.~A.}\ \bibnamefont {Cummer}}, \bibinfo {author}
  {\bibfnamefont {J.~B.}\ \bibnamefont {Pendry}}, \bibinfo {author}
  {\bibfnamefont {A.~F.}\ \bibnamefont {Starr}},\ and\ \bibinfo {author}
  {\bibfnamefont {D.~R.}\ \bibnamefont {Smith}},\ }\href@noop {} {\bibfield
  {journal} {\bibinfo  {journal} {Science}\ }\textbf {\bibinfo {volume}
  {314}},\ \bibinfo {pages} {977} (\bibinfo {year} {2006})}\BibitemShut
  {NoStop}%
\bibitem [{\citenamefont {Nguyen}(2017)}]{Nguyen2017}%
  \BibitemOpen
  \bibfield  {author} {\bibinfo {author} {\bibfnamefont {H.-M.}\ \bibnamefont
  {Nguyen}},\ }\href@noop {} {\bibfield  {journal} {\bibinfo  {journal} {SIAM
  J. Math. Anal.}\ }\textbf {\bibinfo {volume} {49}},\ \bibinfo {pages} {3208}
  (\bibinfo {year} {2017})}\BibitemShut {NoStop}%
\bibitem [{\citenamefont {Guenneau}\ \emph {et~al.}(2005)\citenamefont
  {Guenneau}, \citenamefont {Gralak},\ and\ \citenamefont
  {Pendry}}]{guenneau2005perfect}%
  \BibitemOpen
  \bibfield  {author} {\bibinfo {author} {\bibfnamefont {S.}~\bibnamefont
  {Guenneau}}, \bibinfo {author} {\bibfnamefont {B.}~\bibnamefont {Gralak}},\
  and\ \bibinfo {author} {\bibfnamefont {J.}~\bibnamefont {Pendry}},\
  }\href@noop {} {\bibfield  {journal} {\bibinfo  {journal} {Opt. Lett.}\
  }\textbf {\bibinfo {volume} {30}},\ \bibinfo {pages} {1204} (\bibinfo {year}
  {2005})}\BibitemShut {NoStop}%
\bibitem [{\citenamefont {Guenneau}\ \emph {et~al.}(2010)\citenamefont
  {Guenneau}, \citenamefont {Farhat},\ and\ \citenamefont
  {Enoch}}]{guenneau2010perfect}%
  \BibitemOpen
  \bibfield  {author} {\bibinfo {author} {\bibfnamefont {S.}~\bibnamefont
  {Guenneau}}, \bibinfo {author} {\bibfnamefont {M.}~\bibnamefont {Farhat}},\
  and\ \bibinfo {author} {\bibfnamefont {S.}~\bibnamefont {Enoch}},\
  }\href@noop {} {\bibfield  {journal} {\bibinfo  {journal} {Physica B:
  Condensed Matter}\ }\textbf {\bibinfo {volume} {405}},\ \bibinfo {pages}
  {2947} (\bibinfo {year} {2010})}\BibitemShut {NoStop}%
\bibitem [{\citenamefont {de~Rosny}\ \emph {et~al.}(2000)\citenamefont
  {de~Rosny}, \citenamefont {Tourin},\ and\ \citenamefont {Fink}}]{Rosny2000}%
  \BibitemOpen
  \bibfield  {author} {\bibinfo {author} {\bibfnamefont {J.}~\bibnamefont
  {de~Rosny}}, \bibinfo {author} {\bibfnamefont {A.}~\bibnamefont {Tourin}},\
  and\ \bibinfo {author} {\bibfnamefont {M.}~\bibnamefont {Fink}},\ }\href@noop
  {} {\bibfield  {journal} {\bibinfo  {journal} {Phys. Rev. Lett.}\ }\textbf
  {\bibinfo {volume} {84}},\ \bibinfo {pages} {1693} (\bibinfo {year}
  {2000})}\BibitemShut {NoStop}%
\bibitem [{\citenamefont {Catheline}\ \emph {et~al.}(2011)\citenamefont
  {Catheline}, \citenamefont {Gallot}, \citenamefont {Roux}, \citenamefont
  {Ribay},\ and\ \citenamefont {de~Rosny}}]{Catheline}%
  \BibitemOpen
  \bibfield  {author} {\bibinfo {author} {\bibfnamefont {S.}~\bibnamefont
  {Catheline}}, \bibinfo {author} {\bibfnamefont {T.}~\bibnamefont {Gallot}},
  \bibinfo {author} {\bibfnamefont {P.}~\bibnamefont {Roux}}, \bibinfo {author}
  {\bibfnamefont {G.}~\bibnamefont {Ribay}},\ and\ \bibinfo {author}
  {\bibfnamefont {J.}~\bibnamefont {de~Rosny}},\ }\href@noop {} {\bibfield
  {journal} {\bibinfo  {journal} {Wave Motion}\ }\textbf {\bibinfo {volume}
  {48}},\ \bibinfo {pages} {214} (\bibinfo {year} {2011})}\BibitemShut
  {NoStop}%
\bibitem [{\citenamefont {Wolf}\ and\ \citenamefont {Maret}(1985)}]{Wolf1985}%
  \BibitemOpen
  \bibfield  {author} {\bibinfo {author} {\bibfnamefont {P.-E.}\ \bibnamefont
  {Wolf}}\ and\ \bibinfo {author} {\bibfnamefont {G.}~\bibnamefont {Maret}},\
  }\href@noop {} {\bibfield  {journal} {\bibinfo  {journal} {Phys. Rev. Lett.}\
  }\textbf {\bibinfo {volume} {55}},\ \bibinfo {pages} {2696} (\bibinfo {year}
  {1985})}\BibitemShut {NoStop}%
\bibitem [{\citenamefont {Albada}\ and\ \citenamefont
  {Lagendijk}(1985)}]{Albada1985}%
  \BibitemOpen
  \bibfield  {author} {\bibinfo {author} {\bibfnamefont {M.~P.~V.}\
  \bibnamefont {Albada}}\ and\ \bibinfo {author} {\bibfnamefont
  {A.}~\bibnamefont {Lagendijk}},\ }\href@noop {} {\bibfield  {journal}
  {\bibinfo  {journal} {Phys. Rev. Lett.}\ }\textbf {\bibinfo {volume} {55}},\
  \bibinfo {pages} {2692} (\bibinfo {year} {1985})}\BibitemShut {NoStop}%
\bibitem [{\citenamefont {Pendry}(2008)}]{pendry2008time}%
  \BibitemOpen
  \bibfield  {author} {\bibinfo {author} {\bibfnamefont {J.~B.}\ \bibnamefont
  {Pendry}},\ }\href@noop {} {\bibfield  {journal} {\bibinfo  {journal}
  {Science}\ }\textbf {\bibinfo {volume} {322}},\ \bibinfo {pages} {71}
  (\bibinfo {year} {2008})}\BibitemShut {NoStop}%
\bibitem [{\citenamefont {Krishnan}\ \emph {et~al.}(2001)\citenamefont
  {Krishnan}, \citenamefont {Thio}, \citenamefont {Kim}, \citenamefont {Lezec},
  \citenamefont {Ebbesen}, \citenamefont {Wolff}, \citenamefont {Pendry},
  \citenamefont {Martin-Moreno},\ and\ \citenamefont
  {Garcia-Vidal}}]{krishnan2001evanescently}%
  \BibitemOpen
  \bibfield  {author} {\bibinfo {author} {\bibfnamefont {A.}~\bibnamefont
  {Krishnan}}, \bibinfo {author} {\bibfnamefont {T.}~\bibnamefont {Thio}},
  \bibinfo {author} {\bibfnamefont {T.}~\bibnamefont {Kim}}, \bibinfo {author}
  {\bibfnamefont {H.}~\bibnamefont {Lezec}}, \bibinfo {author} {\bibfnamefont
  {T.}~\bibnamefont {Ebbesen}}, \bibinfo {author} {\bibfnamefont
  {P.}~\bibnamefont {Wolff}}, \bibinfo {author} {\bibfnamefont
  {J.}~\bibnamefont {Pendry}}, \bibinfo {author} {\bibfnamefont
  {L.}~\bibnamefont {Martin-Moreno}},\ and\ \bibinfo {author} {\bibfnamefont
  {F.}~\bibnamefont {Garcia-Vidal}},\ }\href@noop {} {\bibfield  {journal}
  {\bibinfo  {journal} {Opt. Commun.}\ }\textbf {\bibinfo {volume} {200}},\
  \bibinfo {pages} {1} (\bibinfo {year} {2001})}\BibitemShut {NoStop}%
\bibitem [{\citenamefont {Xeridat}(2011)}]{xeridat2011etude}%
  \BibitemOpen
  \bibfield  {author} {\bibinfo {author} {\bibfnamefont {O.}~\bibnamefont
  {Xeridat}},\ }\emph {\bibinfo {title} {Etude exp{\'e}rimentale de la
  propagation, de la diffusion et de la localisation des ondes de Lamb}},\
  \href@noop {} {Ph.D. thesis},\ \bibinfo  {school} {Université de Nice}
  (\bibinfo {year} {2011})\BibitemShut {NoStop}%
\bibitem [{\citenamefont {Dubois}(2014)}]{dubois2014controle}%
  \BibitemOpen
  \bibfield  {author} {\bibinfo {author} {\bibfnamefont {M.}~\bibnamefont
  {Dubois}},\ }\emph {\bibinfo {title} {Contr{\^o}le des ondes de flexion dans
  les plaques}},\ \href@noop {} {Ph.D. thesis},\ \bibinfo  {school}
  {Universit{\'e} Paris Diderot-Paris 7-Sorbonne Paris Cit{\'e}} (\bibinfo
  {year} {2014})\BibitemShut {NoStop}%
\bibitem [{\citenamefont {Dubois}\ \emph {et~al.}(2017)\citenamefont {Dubois},
  \citenamefont {Shi}, \citenamefont {Zhu}, \citenamefont {Wang},\ and\
  \citenamefont {Zhang}}]{Dubois2017}%
  \BibitemOpen
  \bibfield  {author} {\bibinfo {author} {\bibfnamefont {M.}~\bibnamefont
  {Dubois}}, \bibinfo {author} {\bibfnamefont {C.}~\bibnamefont {Shi}},
  \bibinfo {author} {\bibfnamefont {X.}~\bibnamefont {Zhu}}, \bibinfo {author}
  {\bibfnamefont {Y.}~\bibnamefont {Wang}},\ and\ \bibinfo {author}
  {\bibfnamefont {X.}~\bibnamefont {Zhang}},\ }\href@noop {} {\bibfield
  {journal} {\bibinfo  {journal} {Nat. Commun.}\ }\textbf {\bibinfo {volume}
  {8}},\ \bibinfo {pages} {14871} (\bibinfo {year} {2017})}\BibitemShut
  {NoStop}%
\bibitem [{\citenamefont {Goodman}(1996)}]{Goodman}%
  \BibitemOpen
  \bibfield  {author} {\bibinfo {author} {\bibfnamefont {J.~W.}\ \bibnamefont
  {Goodman}},\ }\href@noop {} {\emph {\bibinfo {title} {Introduction to Fourier
  optics}}}\ (\bibinfo  {publisher} {Mc Graw Hill, New York},\ \bibinfo {year}
  {1996})\BibitemShut {NoStop}%
\bibitem [{\citenamefont {Wang}\ \emph {et~al.}(2020)\citenamefont {Wang},
  \citenamefont {Bargiel}, \citenamefont {Lardet-Vieudrin}, \citenamefont
  {Wang}, \citenamefont {Wang},\ and\ \citenamefont {Laude}}]{Wang2020}%
  \BibitemOpen
  \bibfield  {author} {\bibinfo {author} {\bibfnamefont {T.-T.}\ \bibnamefont
  {Wang}}, \bibinfo {author} {\bibfnamefont {S.}~\bibnamefont {Bargiel}},
  \bibinfo {author} {\bibfnamefont {F.}~\bibnamefont {Lardet-Vieudrin}},
  \bibinfo {author} {\bibfnamefont {Y.-F.}\ \bibnamefont {Wang}}, \bibinfo
  {author} {\bibfnamefont {Y.-S.}\ \bibnamefont {Wang}},\ and\ \bibinfo
  {author} {\bibfnamefont {V.}~\bibnamefont {Laude}},\ }\href@noop {}
  {\bibfield  {journal} {\bibinfo  {journal} {Phys. Rev. Appl.}\ }\textbf
  {\bibinfo {volume} {13}},\ \bibinfo {pages} {014022} (\bibinfo {year}
  {2020})}\BibitemShut {NoStop}%
\bibitem [{\citenamefont {Prada}\ \emph {et~al.}(2009)\citenamefont {Prada},
  \citenamefont {Clorennec}, \citenamefont {Murray},\ and\ \citenamefont
  {Royer}}]{Prada2009}%
  \BibitemOpen
  \bibfield  {author} {\bibinfo {author} {\bibfnamefont {C.}~\bibnamefont
  {Prada}}, \bibinfo {author} {\bibfnamefont {D.}~\bibnamefont {Clorennec}},
  \bibinfo {author} {\bibfnamefont {T.~W.}\ \bibnamefont {Murray}},\ and\
  \bibinfo {author} {\bibfnamefont {D.}~\bibnamefont {Royer}},\ }\href@noop {}
  {\bibfield  {journal} {\bibinfo  {journal} {J. Acoust. Soc. Am.}\ }\textbf
  {\bibinfo {volume} {126}},\ \bibinfo {pages} {620} (\bibinfo {year}
  {2009})}\BibitemShut {NoStop}%
\bibitem [{\citenamefont {Grünsteidl}\ \emph {et~al.}(2018)\citenamefont
  {Grünsteidl}, \citenamefont {Berer}, \citenamefont {Hettich},\ and\
  \citenamefont {Veres}}]{Gruensteidl2018}%
  \BibitemOpen
  \bibfield  {author} {\bibinfo {author} {\bibfnamefont {C.}~\bibnamefont
  {Grünsteidl}}, \bibinfo {author} {\bibfnamefont {T.}~\bibnamefont {Berer}},
  \bibinfo {author} {\bibfnamefont {M.}~\bibnamefont {Hettich}},\ and\ \bibinfo
  {author} {\bibfnamefont {I.}~\bibnamefont {Veres}},\ }\href@noop {}
  {\bibfield  {journal} {\bibinfo  {journal} {Appl. Phys. Lett.}\ }\textbf
  {\bibinfo {volume} {112}},\ \bibinfo {pages} {251905} (\bibinfo {year}
  {2018})}\BibitemShut {NoStop}%
\bibitem [{\citenamefont {Pendry}(2006)}]{Pendry2006}%
  \BibitemOpen
  \bibfield  {author} {\bibinfo {author} {\bibfnamefont {J.~B.}\ \bibnamefont
  {Pendry}},\ }\href@noop {} {\bibfield  {journal} {\bibinfo  {journal}
  {Science}\ }\textbf {\bibinfo {volume} {312}},\ \bibinfo {pages} {1780}
  (\bibinfo {year} {2006})}\BibitemShut {NoStop}%
\bibitem [{\citenamefont {Pendry}\ \emph {et~al.}(2012)\citenamefont {Pendry},
  \citenamefont {Aubry}, \citenamefont {Smith},\ and\ \citenamefont
  {Maier}}]{Pendry2012}%
  \BibitemOpen
  \bibfield  {author} {\bibinfo {author} {\bibfnamefont {J.~B.}\ \bibnamefont
  {Pendry}}, \bibinfo {author} {\bibfnamefont {A.}~\bibnamefont {Aubry}},
  \bibinfo {author} {\bibfnamefont {D.~R.}\ \bibnamefont {Smith}},\ and\
  \bibinfo {author} {\bibfnamefont {S.~A.}\ \bibnamefont {Maier}},\ }\href@noop
  {} {\bibfield  {journal} {\bibinfo  {journal} {Science}\ }\textbf {\bibinfo
  {volume} {337}},\ \bibinfo {pages} {549} (\bibinfo {year}
  {2012})}\BibitemShut {NoStop}%
\bibitem [{\citenamefont {Lefebvre}\ \emph {et~al.}(2015)\citenamefont
  {Lefebvre}, \citenamefont {Dubois}, \citenamefont {Beauvais}, \citenamefont
  {Achaoui}, \citenamefont {Ing}, \citenamefont {Guenneau},\ and\ \citenamefont
  {Sebbah}}]{lefebvre2015experiments}%
  \BibitemOpen
  \bibfield  {author} {\bibinfo {author} {\bibfnamefont {G.}~\bibnamefont
  {Lefebvre}}, \bibinfo {author} {\bibfnamefont {M.}~\bibnamefont {Dubois}},
  \bibinfo {author} {\bibfnamefont {R.}~\bibnamefont {Beauvais}}, \bibinfo
  {author} {\bibfnamefont {Y.}~\bibnamefont {Achaoui}}, \bibinfo {author}
  {\bibfnamefont {R.~K.}\ \bibnamefont {Ing}}, \bibinfo {author} {\bibfnamefont
  {S.}~\bibnamefont {Guenneau}},\ and\ \bibinfo {author} {\bibfnamefont
  {P.}~\bibnamefont {Sebbah}},\ }\href@noop {} {\bibfield  {journal} {\bibinfo
  {journal} {Appl. Phys. Lett.}\ }\textbf {\bibinfo {volume} {106}},\ \bibinfo
  {pages} {024101} (\bibinfo {year} {2015})}\BibitemShut {NoStop}%
\bibitem [{\citenamefont {Tian}\ and\ \citenamefont {Yu}(2017)}]{Tian2017}%
  \BibitemOpen
  \bibfield  {author} {\bibinfo {author} {\bibfnamefont {Z.}~\bibnamefont
  {Tian}}\ and\ \bibinfo {author} {\bibfnamefont {L.}~\bibnamefont {Yu}},\
  }\href@noop {} {\bibfield  {journal} {\bibinfo  {journal} {J. Appl. Phys.}\
  }\textbf {\bibinfo {volume} {122}},\ \bibinfo {pages} {234902} (\bibinfo
  {year} {2017})}\BibitemShut {NoStop}%
\bibitem [{\citenamefont {Tang}\ \emph {et~al.}(2020)\citenamefont {Tang},
  \citenamefont {Xu}, \citenamefont {Guenneau},\ and\ \citenamefont
  {Sebbah}}]{Tang2020}%
  \BibitemOpen
  \bibfield  {author} {\bibinfo {author} {\bibfnamefont {K.}~\bibnamefont
  {Tang}}, \bibinfo {author} {\bibfnamefont {C.}~\bibnamefont {Xu}}, \bibinfo
  {author} {\bibfnamefont {S.}~\bibnamefont {Guenneau}},\ and\ \bibinfo
  {author} {\bibfnamefont {P.}~\bibnamefont {Sebbah}},\ }\href@noop {}
  {\bibfield  {journal} {\bibinfo  {journal} {arXiv: 2008.12402}\ } (\bibinfo
  {year} {2020})}\BibitemShut {NoStop}%
\bibitem [{\citenamefont {Zhao}\ and\ \citenamefont {Yu}(2020)}]{Zhao2020}%
  \BibitemOpen
  \bibfield  {author} {\bibinfo {author} {\bibfnamefont {L.}~\bibnamefont
  {Zhao}}\ and\ \bibinfo {author} {\bibfnamefont {M.}~\bibnamefont {Yu}},\
  }\href@noop {} {\bibfield  {journal} {\bibinfo  {journal} {Sci. Rep.}\
  }\textbf {\bibinfo {volume} {10}},\ \bibinfo {pages} {14556} (\bibinfo {year}
  {2020})}\BibitemShut {NoStop}%
\bibitem [{\citenamefont {Guenneau}\ \emph {et~al.}(2021)\citenamefont
  {Guenneau}, \citenamefont {Lombard},\ and\ \citenamefont
  {Bellis}}]{Guenneau2021}%
  \BibitemOpen
  \bibfield  {author} {\bibinfo {author} {\bibfnamefont {S.}~\bibnamefont
  {Guenneau}}, \bibinfo {author} {\bibfnamefont {B.}~\bibnamefont {Lombard}},\
  and\ \bibinfo {author} {\bibfnamefont {C.}~\bibnamefont {Bellis}},\
  }\href@noop {} {\bibfield  {journal} {\bibinfo  {journal} {Appl. Phys.
  Lett.}\ }\textbf {\bibinfo {volume} {118}},\ \bibinfo {pages} {191102}
  (\bibinfo {year} {2021})}\BibitemShut {NoStop}%
\bibitem [{\citenamefont {Li}\ and\ \citenamefont {Chan}(2004)}]{Li2004}%
  \BibitemOpen
  \bibfield  {author} {\bibinfo {author} {\bibfnamefont {J.}~\bibnamefont
  {Li}}\ and\ \bibinfo {author} {\bibfnamefont {C.~T.}\ \bibnamefont {Chan}},\
  }\href@noop {} {\bibfield  {journal} {\bibinfo  {journal} {Phys. Rev. E}\
  }\textbf {\bibinfo {volume} {70}},\ \bibinfo {pages} {055602} (\bibinfo
  {year} {2004})}\BibitemShut {NoStop}%
\bibitem [{\citenamefont {Ambati}\ \emph {et~al.}(2007)\citenamefont {Ambati},
  \citenamefont {Fang}, \citenamefont {Sun},\ and\ \citenamefont
  {Zhang}}]{Ambati2007}%
  \BibitemOpen
  \bibfield  {author} {\bibinfo {author} {\bibfnamefont {M.}~\bibnamefont
  {Ambati}}, \bibinfo {author} {\bibfnamefont {N.}~\bibnamefont {Fang}},
  \bibinfo {author} {\bibfnamefont {C.}~\bibnamefont {Sun}},\ and\ \bibinfo
  {author} {\bibfnamefont {X.}~\bibnamefont {Zhang}},\ }\href@noop {}
  {\bibfield  {journal} {\bibinfo  {journal} {Phys. Rev. B}\ }\textbf {\bibinfo
  {volume} {75}},\ \bibinfo {pages} {195447} (\bibinfo {year}
  {2007})}\BibitemShut {NoStop}%
\bibitem [{\citenamefont {Nguyen}(2015)}]{Nguyen2015}%
  \BibitemOpen
  \bibfield  {author} {\bibinfo {author} {\bibfnamefont {H.-M.}\ \bibnamefont
  {Nguyen}},\ }\href@noop {} {\bibfield  {journal} {\bibinfo  {journal} {Ann.
  Inst. H. Poincaré Anal. NonLin\'{e}aire}\ }\textbf {\bibinfo {volume}
  {32}},\ \bibinfo {pages} {471} (\bibinfo {year} {2015})}\BibitemShut
  {NoStop}%
\end{thebibliography}

\vspace{5 mm}
\newpage

\noindent \textbf{Author contributions}\\

\alex{A.A. initiated and supervised the project. B.G., J.L., C.P. and A.A. designed the complementary devices. Fr.L., Fa.L., C.P. and A.A. designed the elastic superlens. Fr.L. and B.G. performed the numerical simulations. Fr.L., B.G., F.B. and J.L. performed the experiments. All authors analyzed and discussed the results. Fr.L., C.P. and A.A. prepared the manuscript. All authors contributed to finalizing the manuscript.}

\vspace{5mm}
\noindent \textbf{Acknowledgements}\\

The authors thank Wei Guo and Samuel Métais for their works on the numerical aspects of the anti-oject and perfect corner devices, respectively, during their internship.
The authors are also grateful for funding provided by the Agence Nationale de la Recherche (No. ANR-15-CE24-0014-01, Research Project COPPOLA) and by LABEX WIFI (Laboratory of Excellence within the French Program Investments for the Future, No. ANR-10-LABX-24 and No. ANR-10-IDEX-0001-02 PSL*). B.G. acknowledges financial support from the French Direction Générale de l’Armement (DGA).

\end{document}